\newif\ifsubmission
\submissiontrue 

\newif\iffull
\fulltrue 

\newif\ifanonymous
\iffull
  \anonymousfalse
\else
  \anonymoustrue
\fi

\documentclass[sigconf]{acmart}
\renewcommand\footnotetextcopyrightpermission[1]{} 
\usepackage{colortbl}
\usepackage{endnotes}
\usepackage{booktabs}
\usepackage{graphicx}
\usepackage{amsmath}
\usepackage{amsfonts}
\usepackage{amssymb}
\usepackage{multirow}
\usepackage{url}
\usepackage{algorithmic}
\usepackage{algorithm}
\usepackage{tikz}
\usepackage{tikzscale}
\usepackage{cancel}
\usepackage{todonotes}
\usepackage{color}
\usepackage{enumitem}
\usepackage{listings}
\usepackage{array}
\usepackage{footnote}
\usepackage{numprint}
\usepackage{wrapfig}
\usepackage{ragged2e}
\usepackage{lipsum}
\usepackage{cleveref}
\usepackage[binary-units=true, per-mode=symbol, per-symbol=/]{siunitx}
\usepackage[multiple]{footmisc}
\usepackage{csquotes}

\usepackage{pifont}
\newcommand{\cmark}{\ding{51}}
\newcommand{\xmark}{\ding{55}}

\newcolumntype{C}[1]{>{\centering\arraybackslash}p{#1}}

\makesavenoteenv{tabular}
\npthousandsep{,}\npthousandthpartsep{}\npdecimalsign{.}
\graphicspath{{figs/}} 

\newcolumntype{L}[1]{>{\raggedright\let\newline\\\arraybackslash\hspace{0pt}}m{#1}}
\newcolumntype{R}[1]{>{\raggedleft\let\newline\\\arraybackslash\hspace{0pt}}m{#1}}

\makeatletter
\crefformat{section}{\S#2#1#3} 
\crefformat{subsection}{\S#2#1#3}
\crefformat{subsubsection}{\S#2#1#3}

\ifsubmission
  \newcommand{\TODO}[1]{}
  \newcommand{\CHANGED}[1]{#1}
\else 
  \newcommand{\TODO}[1]{\textcolor{red}{TODO: #1}}
  \newcommand{\CHANGED}[1]{\textcolor{blue}{#1}}
\fi

\newcommand{\remove}[1]{}

\newcommand{\sect}[1]{\cref{#1}}

\iffull
	\newcommand{\fig}[1]{Figure~\ref{#1}}
	\newcommand{\tab}[1]{Table~\ref{#1}}

\else 
	\newcommand{\fig}[1]{Fig.~\ref{#1}}
	\newcommand{\tab}[1]{Tab.~\ref{#1}}

\fi

\newcommand{\sys}[1]{Chameleon}

\newcommand\mc[1]{\multicolumn{1}{c}{#1}} 

\newcommand{\mypm}{\mathbin{\mathpalette\@mypm\relax}}
\newcommand{\@mypm}[2]{\ooalign{%
  \raisebox{.1\height}{$#1+$}\cr
  \smash{\raisebox{-.6\height}{$#1-$}}\cr}}

\makeatother

\setcopyright{none}

\acmDOI{}

\acmISBN{}

\begin{document}
\title[\sys{}: A Hybrid Secure Computation Framework for Machine Learning Applications]{\sys{}: A Hybrid Secure Computation Framework \\ for Machine Learning Applications}

\ifanonymous
\author{}
\else
	\author{M. Sadegh Riazi}
	\affiliation{%
		\institution{UC San Diego}
	}
	\email{mriazi@eng.ucsd.edu}

	\author{Christian Weinert}
	\affiliation{%
		\institution{TU Darmstadt, Germany}
	}
	\email{christian.weinert@crisp-da.de}
	
	\author{Oleksandr Tkachenko}
	\affiliation{%
	  \institution{TU Darmstadt, Germany}
	}
	\email{oleksandr.tkachenko@crisp-da.de}
	
	\author{Ebrahim M. Songhori}
	\affiliation{%
		\institution{UC San Diego}
	}
	\email{e.songhori@gmail.com}

	\author{Thomas Schneider}
	\affiliation{%
		\institution{TU Darmstadt, Germany}
	}
	\email{thomas.schneider@crisp-da.de}

	\author{Farinaz Koushanfar}
	\affiliation{%
		\institution{UC San Diego}
	}
	\email{fkoushanfar@eng.ucsd.edu}
\fi

\begin{abstract}
We present \sys{}, a novel hybrid (mixed-protocol) framework for secure function evaluation (SFE) which enables two parties to jointly compute a function without disclosing their private inputs. \sys{} combines the best aspects of generic SFE protocols with the ones that are based upon additive secret sharing. In particular, the framework performs linear operations in the ring $\mathbb{Z}_{2^l}$ using additively secret shared values and nonlinear operations using Yao's Garbled Circuits or the Goldreich-Micali-Wigderson protocol. \sys{} departs from the common assumption of additive or linear secret sharing models where three or more parties need to communicate in the online phase: the framework allows two parties with private inputs to communicate in the online phase under the assumption of a third node generating correlated randomness in an offline phase. Almost all of the heavy cryptographic operations are precomputed in an offline phase which substantially reduces the communication overhead.
\sys{} is both scalable and significantly more efficient than the ABY framework (NDSS'15) it is based on. Our framework supports signed fixed-point numbers. In particular, \sys{}'s vector dot product of signed fixed-point numbers improves the efficiency of mining and classification of encrypted data for algorithms based upon heavy matrix multiplications. Our evaluation of \sys{} on a 5 layer convolutional deep neural network shows 133x and 4.2x faster executions than Microsoft CryptoNets (ICML'16) and MiniONN (CCS'17), respectively.
\end{abstract}



\maketitle



\section{Introduction}\label{sec:intro}
Secure Function Evaluation (SFE) is one of the great achievements of modern cryptography. It allows two or more parties to evaluate a function on their inputs without disclosing their inputs to each other; that is, all inputs are kept private by the respective owners. In fact, SFE emulates a {\it trusted} third party which collects inputs from different parties and returns the result of the target function to all (or a specific set of) parties. There are many applications in privacy-preserving biometric authentication~\cite{DBLP:conf/pet/ErkinFGKLT09,DBLP:conf/icisc/SadeghiSW09,DBLP:conf/sp/OsadchyPJM10,evans2011efficient,blanton2011secure}, secure auctions~\cite{feigenbaum2004secure}, privacy-preserving machine learning~\cite{dowlin2016cryptonets}, and data mining~\cite{DBLP:conf/crypto/LindellP00,DBLP:journals/joc/LindellP02,nikolaenko2013privacy}. 
In 1986, Yao introduced a {\it generic} protocol for SFE, called Yao's Garbled Circuit (GC) protocol~\cite{yao1986generate}. The Goldreich-Micali-Wigderson (GMW) protocol~\cite{goldreich1987play} is another SFE protocol that was introduced in 1987. 

In theory, any function that can be represented as a Boolean circuit can securely be evaluated using GC or GMW protocols. 
However, GC and GMW can often be too slow and hence of limited practical value because they need several symmetric key operations for each gate in the circuit. 
During the past three decades, the great effort of the secure computation community has decreased the overhead of SFE protocols by several orders of magnitude. The innovations and optimizations span the full range from protocol-level to algorithm-level to engineering-level. 
As a result, several frameworks have been designed with the goal of efficiently realizing one (or multiple) SFE protocols. 
They vary by the online/offline run-time, the number of computing nodes (two-party or multi-party), online/offline communication, the set of supported instructions, and the programming language that describes the functionality. 
These frameworks accept the description of the function as either (i) their own customized languages~\cite{malkhi2004fairplay,mood2016frigate}, (ii) high-level languages such as \texttt{C/C++}~\cite{holzer2012secure} or \texttt{Java}~\cite{huang2011faster,liu2015oblivm}, or (iii) Hardware Description Languages (HDLs)~\cite{songhori2015tinygarble,demmler2015automated}.

A number of SFE compilers have been designed for translating a program written in a high level language to low-level code~\cite{malkhi2004fairplay,henecka2010tasty,mood2016frigate}\iffull \cite{BK15,BHWK16,BKJK16}\fi.
The low-level code is supported by other SFE frameworks that serve as a backbone for executing the cryptographic protocols. In addition to generic SFE protocols, additive/linear {\it secret sharing} enables secure computation of linear operations such as multiplication, addition, and subtraction. In general, each framework introduces a set of trade-offs. The frameworks based on secret-sharing require three (or more) computing nodes which operate on distributed shares of variables in parallel and need multiple rounds of communication between nodes to compute an operation on shares of two secret values.

One of the most efficient secure computation frameworks is Sharemind~\cite{bogdanov2008sharemind} which is based on additive secret sharing over the specific ring $\mathbb{Z}_{2^{32}}$. All operations are performed by three computing nodes. Sharemind is secure against honest-but-curious (semi-honest) nodes which are assumed to follow the protocol but they cannot infer any information about the input and intermediate results as long as the majority of nodes are not corrupted. We consider the same adversary model in this paper. Securely computing each operation in Sharemind needs multiple communication rounds between all three nodes which makes the framework relatively slow in the Internet setting. 
Computation based on additive shares in the ring $\mathbb{Z}_{2^{l}}$ enables very efficient and fast linear operations such as Multiplication (MULT), Addition (ADD), and Subtraction (SUB). However, operations such as Comparison (CMP) and Equality test (EQ) are not as efficient and {\it non-linear} operations cannot easily be realized in the ring $\mathbb{Z}_{2^{l}}$. 

We introduce \sys{}, a fast, modular, and hybrid (mixed-protocol) secure two-party computation framework that utilizes GC, GMW, and additive secret sharing protocols and achieves unprecedented performance both in terms of run-time and communication between parties. 
The analogy comes from the fact that similar to a chameleon that changes its color to match the color of the environment, our framework allows changing the executing SFE protocol based on the run-time operation. 
The main design goal behind \sys{} is to create a framework that combines the advantages of the previous secure computation methodologies.
The idea of a mixed-protocol solution was first introduced in \cite{brickell2007privacy} which combines GC with Homomorphic Encryption (HE). HE enables to perform MULT and ADD operations on encrypted values without actually knowing the unencrypted data.

The TASTY framework~\cite{henecka2010tasty} enables automatic generation of protocols based on GC and HE.
However, due to the high computational cost of HE and costly conversion between HE and GC, they achieve only marginal improvement compared to the single protocol execution model \cite{kerschbaum2014automatic}.

Our framework \sys{} is based on ABY~\cite{aby} which implements a hybrid of additive SS, GMW, and GC for efficient realization of SFE.
However, we overcome three major limitations, thereby improving efficiency, scalability, and practicality:
First, ABY's scalability is limited since it only supports combinational circuit descriptions, but most functionalities cannot be efficiently expressed in a combinational-only format~\cite{songhori2015tinygarble}.
Therefore, we add the ability to handle sequential circuits. 
In contrast to combinational circuit representation, sequential circuits are a {\it cyclic} graph of gates and allow for a more compact representation of the functionality.
Second, the ABY model relies on oblivious transfers for precomputing arithmetic triples which we replace by more efficient protocols using a Semi-honest Third Party (STP).
The STP can be a separate computing node or it can be implemented based on a smartcard~\cite{demmler2014ad} or Intel Software Guard Extensions (SGX)~\cite{bahmani2016secure}.
Therefore, the online phase of \sys{} only involves two parties that have private inputs.
Third, we extend ABY to handle signed fixed-point numbers which is needed in many deep learning applications, but not provided by ABY and other state-of-the-art secure computation frameworks such as TASTY.


\sys{} supports 16, 32, and 64 bit signed fixed-point numbers. The number of bits assigned to the fraction and integral part can also be tuned according to the application. 
The input programs to \sys{} can be described in the high-level language \texttt{C++}. The framework itself is also written in \texttt{C++} which delivers fast execution. \sys{} provides a rich library of many non-linear functions such as {\it exp,\ tanh,\ sigmoid,} etc. In addition, the user can simply add any function description as a Boolean circuit or a \texttt{C/C++} program to our framework and use them seamlessly. 

\vspace{0.2in}
{\bf Machine Learning on Private Data Using \sys{}.}
\sys{}'s efficiency helps us to address a major problem in contemporary secure machine learning on private data. 
Matrix multiplication (or equivalently, vector dot product) is one of the most frequent and essential building blocks for many machine learning algorithms and applications. Therefore, in addition to scalability and efficiency described earlier, we design an efficient secure vector dot product protocol based on the Du-Atallah multiplication protocol~\cite{du2001protocols} that has very fast execution and low communication between the two parties. 
We address secure Deep Learning (DL) which is a sophisticated task with increasing attraction. We also provide privacy-preserving classification based on Support Vector Machines (SVMs).

The fact that many pioneering technology companies have started to provide Machine Learning as a Service (MLaaS\CHANGED{\footnote{Amazon AWS AI (\url{https://aws.amazon.com/amazon-ai/})}\footnote{Google Cloud Machine Learning Engine (\url{https://cloud.google.com/ml-engine/})}\footnote{Microsoft Azure Machine Learning Services (\url{https://azure.microsoft.com/services/machine-learning-services/})}}) proves the importance of DL. Deep and Convolutional Neural Networks (DNNs/CNNs) have attracted many machine learning practitioners due to their capabilities and high classification accuracy. In MLaaS, clients provide their inputs to the cloud servers and receive the corresponding results. 
However, the privacy of clients' data is an important driving factor. 
To that end, Microsoft Research has announced CryptoNets~\cite{dowlin2016cryptonets}. CryptoNets is an HE-based methodology that allows secure evaluation (inference) of encrypted queries over {\it already trained} neural networks on the cloud servers. Queries from the clients can be classified securely by the trained neural network model on the cloud server without inferring any information about the query and the result. 
In \sect{ssec:DL}, we show how \sys{} improves over CryptoNets and other previous works.
In addition, we evaluate \sys{} for privacy-preserving classification based on Support Vector Machines (SVMs) in \sect{ssec:SVM}.


\vspace{0.75em}
{\bf Our Contributions.}
In brief, we summarize our main contributions as follows: 
\begin{itemize}
\item We introduce \sys{}, a novel mixed SFE framework based on ABY \cite{aby} which brings benefits upon efficiency, scalability, and practicality by integrating sequential GCs, fixed-point arithmetic, as well as STP-based protocols for precomputing OTs and generating arithmetic and Boolean multiplication triples, and \CHANGED{an optimized} STP-based vector dot product protocol for vector/matrix multiplications. 
\item We provide detailed performance evaluation results of \sys{} compared to the state-of-the-art frameworks. Compared to ABY, \sys{} requires up to 321$\times$ and 256$\times$ less communication for generating arithmetic and Boolean multiplication triples, respectively.
\item We give a proof-of-concept implementation and experimental results on deep and convolutional neural networks. Comparing to the state-of-the-art Microsoft CryptoNets~\cite{dowlin2016cryptonets}, we achieve a \numprint{133}x performance improvement. Comparing to the recent work of \cite{liuoblivious}, we achieve a 4.2x performance improvement using a comparable configuration.
\end{itemize}

\section{Preliminaries}\label{sec:prelim}
In this section, we provide a concise overview of the basic protocols and concepts that are used in the paper.
Intermediate values are kept as shares of a secret. 
In each protocol, secrets are represented differently. We denote a share of value $x$, in secret type $T$, and held by party $i$ as $\langle x\rangle^T_i$.

\subsection{Oblivious Transfer Protocol}\label{ssec:ot}
Oblivious Transfer (OT) is a building block for secure computation protocols.
The OT protocol allows a receiving party $\mathcal{R}$ to obliviously select and receive a message from a set of messages that belong to a sending party $\mathcal{S}$, i.e., without letting $\mathcal{S}$ know what was the selected message.
In 1-out-of-2 OT, $\mathcal{S}$ has two $l$-bit messages ${x_0, x_1}$ and $\mathcal{R}$ has a bit $b$ indicating the index of the desired message.
After performing the protocol, $\mathcal{R}$ obtains $x_b$ without learning anything about $x_{1-b}$ and $\mathcal{S}$ learns no information about $b$. We denote $n$ parallel 1-out-of-2 OTs on $l$-bit messages as $OT_l^n$.

The OT protocol requires costly public-key cryptography that significantly degrades the performance of secure computation.
A number of methods have been proposed to perform a large number of OTs using only a few public-key encryptions together with less costly symmetric key cryptography in a constant number of communication rounds~\cite{beaver1996correlated, ishai2003extending, asharov2013more}.
Although the OT extension methods significantly reduce the cost compared to that of the original OT, the cost is still prohibitively large for complex secure computation that relies heavily on OT.
However, with the presence of a semi-trusted third party, the parties can perform OT protocols with very low cryptographic computation cost as explained in \sect{ssec:fot}.

\subsection{Garbled Circuit Protocol}\label{ssec:gc}
One of the most efficient solutions for generic secure two-party computation is Yao's Garbled Circuit (GC) protocol \cite{yao1986generate} that requires only a constant number of communication rounds.
In the GC protocol, two parties, Alice and Bob, wish to compute a function $f(a,b)$ where $a$ is Alice's private input and $b$ is Bob's.
The function $f(.,.)$ has to be represented as a Boolean circuit consisting of two-input gates, e.g., AND, XOR.
For each wire $w$ in the circuit, Alice generates and assigns two random $k$-bit strings, called \textit{labels}, $X^0_w$ and $X^1_w$ representing 0 and 1 Boolean values where $k$ is a security parameter, usually set to $k=128$ \cite{bellare2013efficient}.
Next, she encrypts the output labels of a gate using the two corresponding input labels as the encryption keys and creates a four-entry table called {\it garbled table} for each gate. 
The garbled table's rows are shuffled according to the point-and-permute technique~\cite{naor1999privacy} where the four rows are permuted by using the Least Significant Bit (LSB) of the input labels as the permutation bits.
Alice sends the garbled tables of all the gates in the circuit to Bob along with the labels corresponding to her input $a$.
Bob also obliviously receives the labels for his inputs from Alice through OT.
He then decrypts the garbled tables one by one to obtain the output labels of the circuit's output wires.
Alice on the other hand has the mapping of the output labels to 0 and 1 Boolean values.
They can learn the output of the function by sharing this information.

\subsection{GMW Protocol}\label{ssec:gmw}
The Goldreich-Micali-Wigderson (GMW) protocol is an interactive secure multi-party computation protocol \cite{goldreich1987play, goldreich2009foundations}.
In the two-party GMW protocol, Alice and Bob compute $f(a, b)$ using the secret-shared values, where $a$ is Alice's private input and $b$ is Bob's.
Similar to the GC protocol, the function $f(.,.)$ has to be represented as a Boolean circuit.
In GMW, the Boolean value of a wire in the circuit is shared between the parties.
Alice has $\langle v \rangle^B_0$ and Bob has $\langle v \rangle^B_1$ and the actual Boolean value is $v = \langle v \rangle^B_0 \oplus \langle v \rangle^B_1$.
Since the XOR operation is associative, the XOR gates in the circuit can be evaluated locally and without any communication between the parties.
The secure evaluation of AND gates requires interaction and communication between the parties.
The communication for the AND gates in the same level of the circuit can be done in parallel.
Suppose an AND gate $x \wedge y = z$ (where $\wedge$ is the AND operation) where Alice has shares $\langle x \rangle^B_0$ and $\langle y \rangle^B_0$, Bob has shares $\langle x \rangle^B_1$ and $\langle y \rangle^B_1$, and they wish to obtain shares $\langle z \rangle^B_0$ and $\langle z \rangle^B_1$ respectively.

As shown in \cite{aby}, the most efficient method for evaluating AND gates in the GMW protocol is based on Beaver's multiplication triples~\cite{beaver1991efficient}: 
%
Multiplication triples are random shared-secrets $a$, $b$, and $c$ such that $\langle c \rangle^B_0 \oplus \langle c \rangle^B_1 = (\langle a \rangle^B_0 \oplus \langle a \rangle^B_1) \wedge (\langle b \rangle^B_0 \oplus \langle b \rangle^B_1)$.
The triples can be generated offline using OTs (cf. \cite{SZ13}) or by a semi-trusted third party (cf.~\sect{ssec:gmt}).
During the online phase, Alice and Bob use the triples to mask and exchange their inputs of the AND gate: $\langle d \rangle^B_i = \langle x \rangle^B_i \oplus \langle a \rangle^B_i$ and $ \langle e \rangle^B_i = \langle y \rangle^B_i \oplus \langle b \rangle^B_i$.
After that, both can reconstruct $d = \langle d \rangle^B_0 \oplus \langle d \rangle^B_1$ and $e = \langle e \rangle^B_0 \oplus \langle e \rangle^B_1$.
This way, the output shares can be computed as $\langle z \rangle^B_0 = (d \wedge e) \oplus (\langle b \rangle^B_0 \wedge d) \oplus (\langle a \rangle^B_0 \wedge e) \oplus \langle c \rangle^B_0$ and $\langle z \rangle^B_1 = (\langle b \rangle^B_1 \wedge d) \oplus (\langle a \rangle^B_1 \wedge e) \oplus \langle c \rangle^B_1$.


\subsection{Additive Secret Sharing}\label{ssec:sss}
In this protocol, a value is shared between two parties such that the addition of two secrets yields the true value. All operations are performed in the ring $\mathbb{Z}_{2^l}$ (integers modulo $2^l$) where each number is represented as an $l$-bit integer. A ring is a set of numbers which is closed under addition and multiplication.

In order to additively share a secret $x$, a random number within the ring is selected, $r\in_R\mathbb{Z}_{2^l}$, and two shares are created as $\langle x\rangle^A_0=r$ and $\langle x\rangle^A_1=x-r\ \text{mod}\ 2^l$. A party that wants to share a secret sends one of the shares to the other party. To reconstruct a secret, one needs to only add two shares $x=\langle x\rangle^A_0+\langle x\rangle^A_1\ \text{mod}\ 2^l$.

Addition, subtraction, and multiplication by a public constant value $\eta$ ($z=x \circ \eta$) can be done locally by the two parties without any communication: party $i$ computes the share of the result as $\langle z\rangle^A_i = \langle x\rangle^A_i \circ \eta\ \text{mod}\ 2^l$, where $\circ$ denotes any of the aforementioned three operations. Adding/subtracting two secrets ($z=x \mypm y$) also does not require any communication and can be realized as $\langle z\rangle^A_i = \langle x\rangle^A_i \mypm \langle y\rangle^A_i\ \text{mod}\ 2^l$. Multiplying two secrets, however, requires one round of communication. Furthermore, the two parties need to have shares of precomputed Multiplication Triples (MT). MTs refer to a set of three shared numbers such that $c=a\times b$. In the offline phase, party $i$ receives $\langle a\rangle^A_i$, $\langle b\rangle^A_i$, and $\langle c\rangle^A_i$ (cf.\ \sect{ssec:gmt}). By having shares of an MT, multiplication is performed as follows: 

\begin{enumerate}
\item Party $i$ computes $\langle e\rangle^A_i=\langle x\rangle^A_i-\langle a\rangle^A_i$ and\\ $\langle f\rangle^A_i=\langle y\rangle^A_i-\langle b\rangle^A_i$. 
\item Both parties communicate to reconstruct $e$ and $f$. 
\item Party $i$ computes its share of the multiplication as \[\langle z\rangle^A_i = f\times \langle a\rangle^A_i+ e\times \langle b\rangle^A_i+\langle c\rangle^A_i+i\times e \times f\]
\end{enumerate}

For more complex operations, the function can be described as an Arithmetic circuit consisting of only addition and multiplication gates where in each step a single gate is processed accordingly.  

\section{The \sys{} Framework}\label{frmwk}
\sys{} comprises of an {\it offline phase} and an {\it online phase}.
The online phase is a two-party execution model that is run between two parties who wish to perform secure computation on their data. In the offline phase, a Semi-honest Third Party (STP) creates {\it correlated randomness} together with random seeds and provides it to the two parties as suggested in~\cite{cr}. We describe how the STP can be implemented in \sect{ssec:shtp} and its role in \sect{ssec:asse}.

The online phase itself consists of three execution environments: GC, GMW, and Additive Secret Sharing (A-SS). We described the functionality of the GC and GMW protocols in \sect{sec:prelim} and we detail our implementations of these protocols in \sect{ssec:gce}. 
We implement two different protocols for the multiplication operation on additive shares: a protocol based on Multiplication Triples (MT) that we described in \sect{ssec:sss} and an optimized version of the Du-Atallah (DA) protocol~\cite{du2001protocols} (cf.\ \sect{ssec:asse}).
In \sect{ssec:oef}, we explain how the online phase works. 
In order to support highly efficient secure computations, all operations that do not depend on the run-time variables are shifted to the offline phase.
The only cryptographic operations in the online phase are the Advanced Encryption Standard (AES) operations that are used in GC for which dedicated hardware acceleration is available in many processors via the AES-NI instruction set. 


The offline phase includes performing four different tasks: (i) precomputing all required OTs that are used in GC and type conversion protocols, thereby providing a very fast {\it encryption-free} online phase for OT, (ii) precomputing Arithmetic Multiplication Triples (A-MT) used in the multiplication of additive secret shares, (iii) precomputing Boolean Multiplication Triples (B-MT) used in the GMW protocol, and lastly, (iv) precomputing vector dot product shares (VDPS) used in the Du-Atallah protocol~\cite{du2001protocols}. 
In order to reduce the communication in the offline phase from the STP to the two parties, we use the seed expansion technique~\cite{demmler2014ad} for generating A-MTs and B-MTs (cf.\ \sect{ssec:gmt}). We also introduce a novel technique that reduces the communication for generating VDPSs (cf.\ \sect{ssec:asse}).

\subsection{\sys{} Online Execution Flow}\label{ssec:oef}
In this section, we provide a high-level description of the execution flow of the online phase. 
As discussed earlier, linear operations such as ADD, SUB, and MULT are executed in A-SS. The dot product of two vectors of size $n$ is also executed in A-SS which comprises $n$ MULTs and $n-1$ ADDs. 
Non-linear operations such as CMP, EQ, MUX and bitwise XOR, AND, OR operations are executed in the GMW or GC protocol depending on which one is more efficient. Recall that in order to execute a function using the GMW or GC protocol, the function has to be described as a Boolean circuit.

However, the most efficient Boolean circuit description of a given function is different for the GMW and the GC protocol:
In the GC protocol, the computation and communication costs only depend on the {\it total number of AND gates} ($N_{\textit{AND}}$) in the circuit. Regardless of the number of XOR gates, functionality, and depth of the circuit, GC executes in a {\it constant} number of rounds. Communication is a linear function of the number of AND gates ($2\times k \times N_{\textit{AND}}$). Due to the Half-Gate optimization (cf. \sect{ssec:gce}), computation is bounded by constructing the garbled tables (four fixed-key AES encryptions) and evaluating them (two fixed-key AES encryptions).
The GMW protocol, on the other hand, has a different computation and communication model. It needs only bit-level AND and XOR operations for the computation but one round of communication is needed per layer of AND gates. Therefore, the most efficient representation of a function in the GMW protocol is the one that has minimum {\it circuit depth}, more precisely, the minimum number of sequentially dependent layers of AND gates. As a result, when the network latency is high or the depth of the circuit is high, we use GC to execute non-linear functions, otherwise, GMW will be utilized. The computation and communication costs for atomic operations are given in~\sect{sec:eval}.


The program execution in \sys{} is described as different layers of operations where each layer is most efficiently realized in one of the execution environments. 
The execution starts from the first layer and the corresponding execution environment. Once all operations in the first layer are finished, \sys{} switches the underlying protocol and continues the process in the second execution environment. Changing the execution environment requires that the type of the shared secrets should be changed in order to enable the second protocol to continue the process.
One necessary condition is that the cost of the share type translation must not be very high to avoid diminishing the efficiency achieved by the hybrid execution. For converting between the different sharing types, we use the methods from the ABY framework~\cite{aby} which are based on highly efficient OT extensions.

{\bf Communication Rounds.} The number of rounds that both parties need to communicate in \sys{} depends on the number of switches between execution environments and the depth of the circuits used in the GMW protocol. 
We want to emphasize that the number of communication rounds does not depend on the size of input data. Therefore, the network latency added to the execution time is quickly amortized over a high volume of input data.

\subsection{Security Model}
\sys{} is secure against honest-but-curious (HbC), a.k.a. semi-honest, adversaries. This is the standard security model in the literature and considers adversaries that follow the protocol but attempt to extract more information based on the data they receive and process. Honest-but-curious is the security model for the great majority of prior art, e.g., Sharemind~\cite{bogdanov2008sharemind}, ABY~\cite{aby}, and TinyGarble~\cite{songhori2015tinygarble}. 

The Semi-honest Third Party (STP) can be either implemented using a physical entity, in a distributed manner using MPC among multiple non-colluding parties, using trusted hardware (hardware security modules or smartcards~\cite{demmler2014ad}), or using trusted execution environments such as Intel SGX~\cite{bahmani2016secure}.
In case the STP is implemented as a separate physical computation node, our framework is secure against semi-honest adversaries with an honest majority. The latter is identical to the security model considered in the Sharemind framework~\cite{bogdanov2008sharemind}. \CHANGED{In \S\ref{sec:rw}, we list further works based on similar assumptions.}
\CHANGED{Please note that we introduce a new and more practical \emph{computational} model that is superior to Sharemind since only two primary parties are involved in the online execution. This results in a significantly faster run-time while better matching real-world requirements.}

\subsection{Semi-honest Third Party (STP)}\label{ssec:shtp}
In \sys{}, the STP is only involved in the offline phase in order to generate correlated randomness~\cite{cr}. It is not involved in the online phase and thus does not receive any information about the two parties' inputs nor the program being executed. The only exception is computing VDPS for the Du-Atallah protocol in which the STP needs to know the size of the vectors in each dot product \CHANGED{beforehand}. 
Since the security model in \sys{} is HbC with honest majority, some information can be revealed if the STP colludes with either party.

\CHANGED{In order to prevent the STP from observing communication between the two parties, authenticated encryption is added to the communication channel.}
\CHANGED{Likewise it is advised to encrypt communication between the STP and the two parties so they cannot reconstruct the opposite party's private inputs from observed and received messages.}

\section{\sys{} Design and Implementation}\label{sec:intro}
In this section, we provide a detailed description of the different components of \sys{}. \sys{} is written in \texttt{C++} and accepts the program written in \texttt{C++}. The implementation of the GC and GMW engines is covered in \sect{ssec:gce} and the A-SS engine is described in \sect{ssec:asse}.
\sect{ssec:ssfpn} illustrates how \sys{} supports signed fixed-point representation. 
The majority of cryptographic operations is shifted from the online phase to the offline phase. Thus, in \sect{ssec:gmt}, we describe the process of generating Arithmetic/Boolean Multiplication Triples (A-MT/B-MT). \sect{ssec:fot} provides our STP-based implementation for fast Oblivious Transfer and finally the security justification of \sys{} is given in \sect{ssec:sj}.

\subsection{GC and GMW Engines}\label{ssec:gce}
\sys{}'s implementation of the GC protocol is based on the methodology presented in~\cite{songhori2015tinygarble}. Therefore, the input to the GC engine is the topologically sorted list of Boolean gates in the circuit as an \texttt{.scd} file. We synthesized GC-optimized circuits and created the \texttt{.scd} files for many primitive functions. A user can simply use these circuits by calling regular functions in the \texttt{C++} language. 
We include most recent optimizations for GC: Free XOR~\cite{kolesnikov2008improved}, fixed-key AES garbling~\cite{bellare2013efficient}, Half Gates~\cite{zahur2015two}, and sequential circuit garbling~\cite{songhori2015tinygarble}.

Our implementation of the GMW protocol is based on the ABY framework~\cite{aby}. Therefore, the function description format of GMW is an \texttt{.aby} file. All the circuits are depth-optimized as described in~\cite{demmler2015automated} to incur the least latency during the protocol execution. \sys{} users can simply use these circuits by calling a function with proper inputs. 

\subsection{A-SS Engine}\label{ssec:asse}
In \sys{}, linear operations, i.e., ADD, SUB, MULT, are performed using additive secret sharing in the ring $\mathbb{Z}_{2^l}$. We discussed in \sect{ssec:sss} how to perform a single MULT using a multiplication triple. However, there are other methods to perform a MULT: (i) The protocol of~\cite{ben2016optimizing} has very low communication in the online phase. However, in contrast to our computation model, it requires STP interaction with the other two parties in the online phase. (ii) The Du-Atallah protocol~\cite{du2001protocols} is another method to perform multiplication on additive shared values which we describe next.

{\bf The Du-Atallah Multiplication Protocol \cite{du2001protocols}. }
In this protocol, two parties $\mathcal{P}_0$ (holding $x$) and $\mathcal{P}_1$ (holding $y$) together with a third party $\mathcal{P}_2$ can perform multiplication $z=x\times y$. At the end of this protocol, $z$ is additively shared between all {\it three} parties. The protocol works as follows: 
\begin{enumerate}
\item $\mathcal{P}_2$ randomly generates $a_0,a_1\in_R \mathbb{Z}_{2^l}$ and sends $a_0$ to $\mathcal{P}_0$ and $a_1$ to $\mathcal{P}_1$.
\item $\mathcal{P}_0$ computes $(x+a_0)$ and sends it to $\mathcal{P}_1$. Similarly, $\mathcal{P}_1$ computes $(y+a_1)$ and sends it to $\mathcal{P}_0$.
\item $\mathcal{P}_0$, $\mathcal{P}_1$, and $\mathcal{P}_2$ can compute their share as $\langle z \rangle^A_0=-a_0\times (y+a_1)$, $\langle z \rangle^A_1=y\times (x+a_0)$, and $\langle z \rangle^A_2=a_0\times a_1$, respectively.
\end{enumerate}
It can be observed that the results are true additive shares of $z$: $\langle z \rangle^A_0+\langle z \rangle^A_1+\langle z \rangle^A_2=z$.
Please note that this protocol computes shares of a multiplication of two numbers held by two parties in {\it cleartext}. In the general case, where both $x$ and $y$ are additively shared between two parties ($\mathcal{P}_0$ holds $\langle x \rangle^A_0$, $\langle y \rangle^A_0$ and $\mathcal{P}_1$ holds $\langle x \rangle^A_1$, $\langle y \rangle^A_1$), the multiplication can be computed as $z=x\times y=(\langle x \rangle^A_0+\langle x \rangle^A_1)\times (\langle y \rangle^A_0+\langle y \rangle^A_1)$. The two terms $\langle x \rangle^A_0 \times \langle y \rangle^A_0$ and $\langle x \rangle^A_1 \times \langle y \rangle^A_1$ can be computed locally by $\mathcal{P}_0$ and $\mathcal{P}_1$, respectively.
{\it Two} instances of the Du-Atallah protocol are needed to compute shares of $\langle x \rangle^A_0 \times \langle y \rangle^A_1$ and $\langle x \rangle^A_1 \times \langle y \rangle^A_0$. Please note that $\mathcal{P}_i$ should not learn $\langle x \rangle^A_{1-i}$ and $\langle y \rangle^A_{1-i}$, otherwise, secret values $x$ and/or $y$ are revealed to $\mathcal{P}_i$.
At the end, $\mathcal{P}_0$ has 
$$\langle x \rangle^A_0 \times \langle y \rangle^A_0,\ \big \langle \langle x \rangle^A_0 \times \langle y \rangle^A_1 \big \rangle^A_0,\ \big \langle \langle x \rangle^A_1 \times \langle y \rangle^A_0\big \rangle^A_0$$ 
and $\mathcal{P}_1$ has 
$$\langle x \rangle^A_1 \times \langle y \rangle^A_1,\ \big \langle \langle x \rangle^A_0 \times \langle y \rangle^A_1 \big \rangle^A_1,\ \big \langle \langle x \rangle^A_1 \times \langle y \rangle^A_0\big \rangle^A_1$$
where $\langle z \rangle^A_0$ and $\langle z \rangle^A_1$ are the summation of each party's shares, respectively. 

The Du-Atallah protocol is used in Sharemind~\cite{bogdanov2008sharemind} where there are three active computing nodes that are involved in the online phase, whereas, in \sys{}, the third party (STP) is only involved in the offline phase. This problem can be solved since the role of $\mathcal{P}_2$ can be shifted to the offline phase as follows: (i) Step one of the Du-Attallah protocol can be computed in the offline phase for as many multiplications as needed. (ii) In addition, $\mathcal{P}_2$ randomly generates another $l$-bit number $a_2$ and computes $a_3=(a_0\times a_1)-a_2$. $\mathcal{P}_2$ sends $a_2$ to $\mathcal{P}_0$ and $a_3$ to $\mathcal{P}_1$ in the offline phase. During the online phase, both parties additionally add their new shares ($a_2$ and $a_3$) to their shared results: $\langle z \rangle^A_{0,\textit{new}}=\langle z \rangle^A_0+a_2$ and $\langle z \rangle^A_{1,\textit{new}}=\langle z \rangle^A_1+a_3$. 
This modification is perfectly secure since $\mathcal{P}_0$ has received a true random number and $\mathcal{P}_1$ has received $a_3$ which is an additive share of $(a_0\times a_1)$. Since $a_2$ has uniform distribution, the probability distribution of $a_3$ is also uniform~\cite{bogdanov2008sharemind} and as a result, $\mathcal{P}_1$ cannot infer additional information. 

{\bf Optimizing the Du-Atallah Protocol. }
As we will discuss in \sect{sec:app}, in many cases, the computation model is such that one operand $x$ is held in cleartext by one party, e.g., $\mathcal{P}_0$, and the other operand $y$ is shared among two parties: $\mathcal{P}_0$ has $\langle y \rangle^A_0$ and $\mathcal{P}_1$ has $\langle y \rangle^A_1$. This situation repeatedly arises when the intermediate result is multiplied by one of the party's inputs which is not shared. In this case, only one instance of the Du-Atallah protocol is needed to compute $x\times\langle y \rangle^A_1$. As analyzed in this section, employing this variant of the Du-Atallah protocol is more efficient than the protocol based on MTs. Please note that in order to utilize MTs, both operands need to be shared among the two parties first, which, as we argue here, is inefficient and unnecessary. \tab{tab:summMP} summarizes the computation and communication costs for the Du-Atallah protocol and the protocol based on MTs (\sect{ssec:sss}). As can be seen, online computation and communication are improved by factor 2x. Also, the offline communication is improved by factor 3x. Unfortunately, using the Du-Atallah protocol in this format will reduce the efficiency of vector dot product computation in \sys{}. Please note that it is no longer possible to perform a complete dot product of two vectors by two parties only since the third share ($\langle z \rangle^A_2=a_0\times a_1$) is shared between two parties ($\mathcal{P}_0$ and $\mathcal{P}_1$). However, this problem can be fixed by a modification which we describe next. 

\begin{table}[]
\centering
\caption{Summary of properties of the Du-Atallah multiplication protocol and the protocol based on Multiplication Triples in \sect{ssec:sss}. ($i,j$) means $\mathcal{P}_0$ and $\mathcal{P}_1$ have to perform $i$ and $j$ multiplications in plaintext, respectively. Offline and online communications are expressed in number of bits. The size of online communication corresponds to data transmission in each direction. $^*$Initial sharing of $x$ is also considered.}
\label{tab:summMP}
\resizebox{\columnwidth}{!}{
\begin{tabular}{|l|l|l|l|l|}
\hline
\textbf{Protocol} & \textbf{\# MULT ops} & \textbf{Online Comm.} & \textbf{Offline Comm.}  & \textbf{Rounds} \\ \hline
Multiplication Triple    & (3,4)  &  $2\cdot l$  & $3\cdot l$ & 2$^*$ \\ \hline
Du-Atallah  & (1,2) & $l$  & $2\cdot l$  & 1 \\ \hline
\end{tabular}}
\end{table} 

{\bf Du-Atallah Protocol and Vector Dot Product. }
We further modify the optimized Du-Atallah protocol such that the complete vector dot product is efficiently processed. The idea is that instead of the STP additively sharing its shares, it first sums its shares and then sends the additively shared versions to the two parties. 
Consider vectors of size $n$. The STP needs to generate $n$ different $a_0$ and $a_1$ as a list for a single vector multiplication. We denote the $j^{\textit{th}}$ member of the list as $[a_0]_j$ and $[a_1]_j$. 
Our modification requires that the STP generates a single $l$-bit value $a_2$ and sends it to $\mathcal{P}_0$. The STP also computes 
$$a_3=\sum_{i=0}^{n-1} [a_0]_j\times [a_1]_j-a_2$$ 
and sends it to $\mathcal{P}_1$. We call $a_2$ and $a_3$ the {\it Vector Dot Product Shares} or VDPS. This requires that the STP knows the size of the array in the offline phase. Since the functionality of the computation is not secret, we can calculate the size and number of all dot products in the offline phase and ask for the corresponding random shares from the STP. 

{\bf Reducing Communication. }
A straightforward implementation of the offline phase of the Du-Atallah protocol requires that the STP sends $\sim n$ random numbers of size $l$ ($[a_0]_j$ and $[a_1]_j$) to $\mathcal{P}_0$ and $\mathcal{P}_1$ for a single dot product of vectors of size $n$. 
However, we suggest reducing the communication using a Pseudo Random Generator (PRG) for generating the random numbers as was proposed in~\cite{demmler2014ad}. Instead of sending the complete list of numbers to each party, the STP can create and send random PRG seeds for each string to the parties such that each party can create $[a_0]_j$ and $[a_1]_j$ locally using the PRG. For this purpose, we implement the PRG using Advanced Encryption Standard (AES), a low-cost block cipher, in counter mode (AES CTR-DRBG). Our implementation follows the description of the NIST Recommendation for DRBG \cite{barker2015nist}. From a 256-bit seed, AES CTR-DRBG can generate $2^{63}$ indistinguishable random bits. If more than $2^{63}$ bits are needed, the STP sends more seeds to the parties.
The STP uses the same seeds in order to generate $a_2$ and $a_3$ for each dot product. Therefore, the communication is reduced from $n\times l$ bits to sending a one-time 256-bit seed and an $l$-bit number per single dot product. 

\CHANGED{{\bf Performance evaluation. }
For an empirical performance evaluation of our optimized VDP protocol, we refer the reader to \sect{ssec:SVM}: the evaluated SVM classification mainly consists of a VDP computation together with a negligible subtraction and comparison operation.
}

\subsection{Supporting Signed Fixed-point Numbers}\label{ssec:ssfpn}
\sys{} supports Signed Fixed-point Numbers (SFN) in addition to integer operations. 
Supporting SFN requires that not only all three secure computation protocols (GC, GMW, and Additive SS) support SFN but the secret translation protocols should be compatible as well. We note that the current version of the ABY framework only supports unsigned integer values. We added an abstraction layer to the ABY framework such that it supports signed fixed-point numbers. The TinyGarble framework can support this type if the corresponding Boolean circuit is created and fed into the framework. 

All additive secret sharing protocols only support unsigned integer values. However, in this section, we describe how such protocols can be modified to support {\it signed fixed-point} numbers. Modification for supporting {\it signed integers} can be done by representing numbers in two's complement format. Consider the ring $\mathbb{Z}_{2^l}$ which consists of unsigned integer numbers $\{0,1,2,...,2^{l-1}-1,2^{l-1},...,2^l-1\}$. We can perform signed operations only by {\it interpreting} these numbers differently as the two's complement format: $\{0,1,2,...,2^{l-1}-1,-2^{l-1},...,-1\}$. By doing so, signed operations work seamlessly.

In order to support fixed-point precision, one solution is to interpret signed integers as signed fixed-point numbers. 
Each number is represented in two's complement format with the Most Significant Bit (MSB) being the sign bit. There are $\alpha$ and $\beta$ bits for integer and fraction parts, respectively. Therefore, the total number of bits is equal to $\gamma = \alpha + \beta +1$. While this works perfectly for addition and subtraction, it cannot be used for multiplication. The reason is that when multiplying two numbers in a ring, the rightmost $2\times \beta$ bits of the result now correspond to the fraction part instead of $\beta$ bits. Also, $\beta$ bits from MSBs are overflown and discarded. 
Our solution to this problem is to perform all operations in the ring $\mathbb{Z}_{2^l}$ where $l=\gamma + \beta$ and after each multiplication, we shift the result $\beta$ bits to the right while replicating the sign bit for $\beta$ MSBs. 

In addition to the support by the computation engines, share translation protocols also work correctly.
Share translation from GC to GMW works fine as it operates on bit-level and is transparent to the number representation format. Share translation from GC/GMW to additive sharing either happens using a subtraction circuit or OT. In the first case, the result is valid since the subtraction of two signed fixed-point numbers in two's complement format is identical to subtracting two unsigned integers. In the second case, OT is on bit-level and again transparent to the representation format. Finally, share translation from additive sharing to GC/GMW is correct because it uses an addition circuit which is identical for unsigned integers and signed fixed-point numbers. 


{\bf Floating Point Operations. }
The current version of \sys{} supports floating point operations by performing all computations in the GC protocol. Since our GC engine is based on TinyGarble, our performance result is identical to that of TinyGarble, hence, we do not report the experimental results of floating-point operations. A future direction of this work can be to break down the primitive floating point operations, e.g., ADD, MULT, SUB, etc. into smaller atomic operations based on integer values. 
Consequently, one can perform the linear operations in the ring and non-linear operations in GC/GMW, providing a faster execution for floating-point operations.

Most methods for secure computation on floating and fixed point numbers proposed in the literature were realized in Shamir's secret sharing scheme, e.g.\ \cite{catrina2010secure, abzs13, zhang2013picco, krips2014hybrid, pullonen2015combining}, but some of them also in GC \cite{pullonen2015combining}, GMW \cite{demmler2015automated}, and HE \cite{liu2016privacy} based schemes. The quality of the algorithms varies from self-made to properly implemented IEEE 754 algorithms, such as in \cite{pullonen2015combining,demmler2015automated}. The corresponding software implementations were done either in the frameworks Sharemind \cite{bogdanov2008sharemind} and PICCO \cite{zhang2013picco}, or as standalone applications. For fixed-point arithmetics, Aliasgari et al.\ \cite{abzs13} proposed algorithms that outperform even integer arithmetic for certain operations. As a future direction of this work, we plan to integrate their methodology in \sys{}.

\subsection{Generating Multiplication Triples}\label{ssec:gmt}
As we discussed in \sect{ssec:sss}, each multiplication on additive secret shares requires an Arithmetic Multiplication Triple (A-MT) and one round of communication. Similarly, evaluating each AND gate in the GMW protocol requires a Boolean Multiplication Triple (B-MT)~\cite{demmler2014ad}. 
In the offline phase, we calculate the number of MTs ($N_\text{A-MT}$ and $N_\text{B-MT}$). The STP precomputes all MTs needed and sends them to both parties. More precisely, to generate A-MTs, the STP uses a PRG to produce five $l$-bit random numbers corresponding to $a_0,b_0,c_0,a_1$, and $b_1$. We denote the $j^{\textit{th}}$ triple with $[.]_{j}$. Therefore, the STP completes MTs by computing $c_1$'s as $[c_1]_j=([a_0]_j+[a_1]_j)\times([b_0]_j+[b_1]_j)-[c_0]_j$. Finally, the STP sends $[a_0]_j, [b_0]_j$, and $[c_0]_j$ to the first party and $[a_1]_j, [b_1]_j$, and $[c_1]_j$ to the second party for $j=1,2,...,N_\text{A-MT}$. Computing B-MTs is also very similar with the only differences that all numbers are 1-bit and $[c_1]_j$ is calculated as $[c_1]_j = ([a_0]_j \oplus [a_1]_j) \wedge ([b_0]_j \oplus [b_1]_j)\oplus [c_0]_j$.

{\bf Reducing Communication. }
A basic implementation of precomputing A-MTs and B-MTs requires communication of $3\times l \times N_\text{A-MT}$ and $3 \times N_\text{B-MT}$ bits from the STP to each party, respectively. However, similar to the idea of \cite{demmler2014ad} presented in \sect{ssec:asse}, we use a PRG to generate random strings from seeds locally by each party.
To summarize the steps:
\begin{enumerate}
\item STP generates two random seeds: $\mathrm{seed}_0$ for generating $[a_0]_j, [b_0]_j$, and $[c_0]_j$ and $\mathrm{seed}_1$ for $[a_1]_j$ and $[b_1]_j$.
\item STP computes $[c_1]_j=([a_0]_j+[a_1]_j)\times([b_0]_j+[b_1]_j)-[c_0]_j$ for $j=1,2,...,N_{A-MT}$.
\item STP sends $\mathrm{seed}_0$ to the first party and $\mathrm{seed}_1$ together with the list of $[c_1]_j$ to the second party.
\end{enumerate}

\begin{figure}[h]
\centering
\includegraphics[width=\columnwidth]{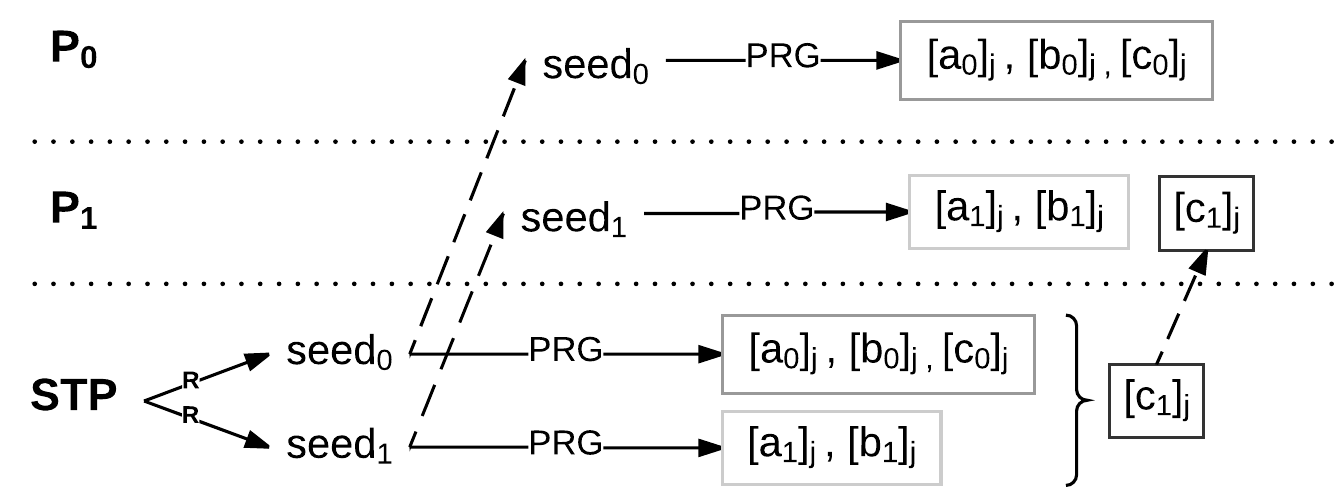}
\caption{\label{fig:seeds}{Seed expansion process to precompute A-MTs/B-MTs with low communication.}}
\end{figure}

After receiving the seeds, both parties locally generate their share of the triples using the same PRG. This method reduces the communication from $3\times l \times N_\text{A-MT}$ to $256$ and $256+l\times N_\text{A-MT}$ bits for the first and second parties, respectively. The STP follows a similar process with the same two seeds to generate B-MTs. \fig{fig:seeds} illustrates the seed expansion idea to generate MTs~\cite{demmler2014ad}.

\subsection{Fast STP-aided Oblivious Transfer}\label{ssec:fot}
\begin{figure}[h]
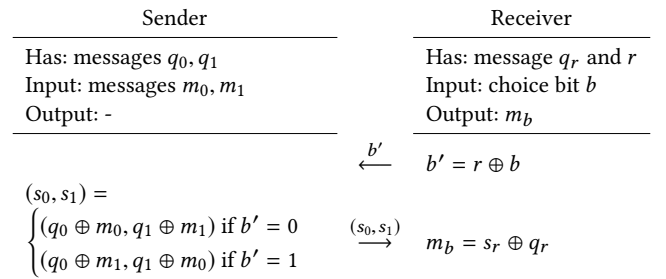

\[ 
\resizebox{\columnwidth}{!}{$\displaystyle
\begin{array}{lcl}
\mc{\text{Sender}} & & \mc{\text{Receiver}} \\
\cmidrule{1-1} \cmidrule{3-3}
\text{Has: messages } q_0, q_1 & &
\text{Has: message } q_r \text{ and } r\\
\text{Input: messages } m_0, m_1 & &
\text{Input: choice bit } b\\
\text{Output: -} & &
\text{Output: } m_b\\
\cmidrule{1-1} \cmidrule{3-3}
 &
  \overset{b'}{\longleftarrow} &
  b' = r \oplus b\\
(s_0,s_1)=\\
\begin{cases}
(q_0 \oplus m_0, q_1 \oplus m_1) \text{ if } b'=0\\
(q_0 \oplus m_1, q_1 \oplus m_0) \text{ if } b'=1
\end{cases} &
\overset{(s_0,s_1)}{\longrightarrow} & m_b = s_r \oplus q_r
  \\
\end{array}
$}
\]
\caption{Beaver's OT precomputation protocol \cite{beaver1995precomputing}.}
\label{p:beaver}
\end{figure}

Utilizing the idea of correlated randomness~\cite{cr}, we present an efficient and fast protocol for Oblivious Transfer that is aided by the Semi-honest Third Party (STP). Our protocol comprises an offline phase (performed by the STP) and an online phase (performed by the two parties).
The protocol is described for one 1-out-of-2 OT. The process repeats for as many OTs as required.
In the offline phase, the STP generates random masks $q_0$, $q_1$ and a random bit $r$ and sends $q_0$, $q_1$ to the sender and $r$, $q_r$ to the receiver.
In the online phase, the two parties execute the online phase of Beaver's OT precomputation protocol~\cite{beaver1995precomputing} described in \fig{p:beaver}.
Please note that all OTs in \sys{} including OTs used in GC and secret translation from GC/GMW to Additive are implemented as described above.

{\bf Reducing Communication. }
Similar to the idea discussed in \sect{ssec:gmt}, the STP does not actually need to send the list of $(q_0, q_1)$ to the sender and $r$ to the receiver. Instead, it generates two random seeds and sends them to the two parties. The STP only needs to send the full list of $q_r$ to the receiver.

\subsection{Security Justification}\label{ssec:sj}
The security proof of \sys{} is based on the following propositions: 
(i) the GC execution is secure since it is based on~\cite{songhori2015tinygarble}. (ii) The security proof of GMW execution and share type translation directly follows the one of~\cite{aby}. (iii) All operations in A-SS are performed in the ring $\mathbb{Z}_{2^l}$ which is proven to be secure in~\cite{bogdanov2008sharemind}. Our support for SFN only involves the utilization of a bigger ring and does not change the security guarantees, and finally (iv) our optimizations for reducing the communication between the STP and the two parties are secure as we use a PRG instantiation recommended by the NIST standard~\cite{barker2015nist}. 

\section{Machine Learning Applications}\label{sec:app}
Many applications can benefit from our framework. Here, we cover only two important applications in greater detail due to the space constraints. In particular, we show how Chameleon can be leveraged in Deep Learning (\sect{ssec:DL}) and classification based on Support Vector Machines (\sect{ssec:SVM}).


\CHANGED{
We run all our experiments for a long-term security parameter (128-bit security) on machines equipped with Intel Core i7-4790 CPUs @ \SI{3.6}{\giga\hertz} and \SI{16}{\giga\byte} of RAM.
The CPUs support fast AES evaluations due to AES-NI.
\CHANGED{The STP is instantiated as a separate compute node running a \texttt{C/C++} implementation.}
The communication between the STP and its clients \CHANGED{as well as between the clients} is \CHANGED{protected by TLS with client authentication}.
Except when stated otherwise, all parties run on different machines within the same Gigabit network.
}

\subsection{Deep Learning}\label{ssec:DL}
We evaluate our framework on Deep Neural Networks (DNNs) and a more sophisticated variant, Convolutional Deep Neural Networks (CNNs). Processing both, DNNs and CNNs, requires the support for signed fixed-point numbers. We compare our results with the state-of-the-art Microsoft CryptoNets~\cite{dowlin2016cryptonets}, which is a customized solution for this purpose based on homomorphic encryption, as well as other recent solutions.

\begin{figure*}[ht]
\centering
\includegraphics[width=\textwidth]{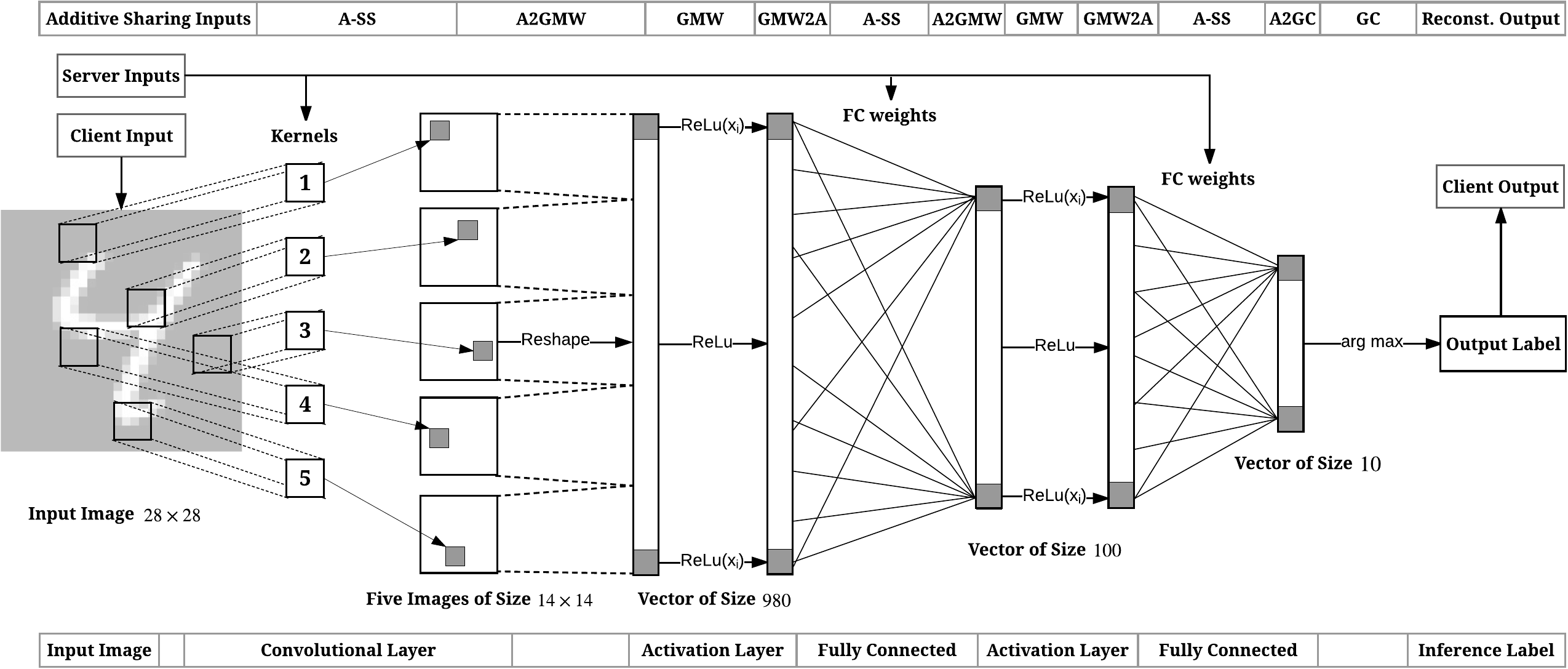}
\caption{\label{fig:arch}{Architecture of our Convolutional Neural Network trained for the MNIST dataset. The upper bar illustrates which protocol is being executed at each phase of the CNN. The lower bar shows different layers of the CNN from the DL perspective.}}
\end{figure*}

{\bf Deep Neural Networks.}
Deep learning is a very powerful method for modeling and classifying raw data that has gained a lot of attention in the past decade due to its superb accuracy. Deep Learning automatically learns complex features using artificial neural networks. 
While there are many different DNNs and CNNs, they all share a similar structure. They are networks of multiple layers stacked on top of each other where the output of each layer is the input to the next layer. 
The input to DNNs is a feature vector, which we denote as ${\bf x}$. The input is passed through the intermediate layers (hidden layers). The output vector of the $L^{\textit{th}}$ layer is ${\bf x}^{(L)}$ where $x_i^{(L)}$ denotes the $i^{\textit{th}}$ element. The length of the vector can change after each layer. The length of the intermediate result vector at layer $L$ is $N_L=\mathrm{length}({\bf x}^{(L)})$. A DNN is composed of a series of (i) {\it Fully Connected layer (FC)}: the output ${\bf x}^{(L)}$ is the matrix multiplication of input vector ${\bf x}^{(L-1)}$ and a matrix weight ${\bf W}$, that is, ${\bf x}^{(L)}={\bf x}^{(L-1)} \cdot {\bf W}^{(L)}$. In general, the size of the input and output of the $FC$ layer is denoted as $FC^{N_{L-1}\times N_L}$. (ii) {\it Activation layer (Act)}: which applies an activation function $f(.)$ on the input vector: ${x}^{(L)}_i=f({x}_i^{(L-1)})$. The activation function is usually a Rectified Linear Unit (ReLu), Tangent-hyperbolic (Tanh), or Sigmoid function \cite{shokri2015privacy,dowlin2016cryptonets}.

The input to a CNN is a picture represented as a matrix ${\bf X}$ where each element corresponds to the value of a pixel. Pictures can have multiple color channels, e.g., RGB, in which case the picture is represented as a multidimensional matrix, a.k.a, {\it tensor}.
CNNs are similar to DNNs but they can potentially have additional layers: (i) {\it Convolution layer (C)} which is essentially a weighted sum of a \enquote{square region} of size $s_q$ in the proceeding layer. To compute the next output, the multiplication window on the input matrix is moved by a specific number, called stride ($s_t$). The weight matrix is called {\it kernel}. There can be $N_{map}$ (called map count) kernels in the convolution layer.  (ii) {\it Mean-pooling (MeP)} which is the average of each square region of the proceeding layer. (iii) {\it Max-pooling (MaP)} is the maximum of each square region of the proceeding layer. The details of all layers are provided in \tab{tab:layers}.

Many giant technology companies such as Google, Microsoft, Facebook, and Apple have invested millions of dollars in accurately training neural networks to serve in different services. Clients that want to use these services currently need to reveal their inputs that may contain sensitive information to the cloud servers. Therefore, there is a special need to run a neural network (trained by the cloud server) on input from another party (clients) while keeping both, the network parameters and the input, private to the respective owners. For this purpose, 
Microsoft has announced CryptoNets~\cite{dowlin2016cryptonets} that can process encrypted queries in neural networks using homomorphic encryption. Next, we compare the performance result of \sys{} to CryptoNets and other more recent works. 

\begin{table}[h]
\centering
\caption{Different types of layers in DNNs and CNNs.}
\label{tab:layers}
\scalebox{0.9}{
\begin{tabular}{|l|l|}
\hline
\textbf{Layer} & \textbf{Functionality} \\ \hline
FC             & $x_{i}^{(L)} = \sum_{j=0}^{N_{L-1}-1} W_{ij}^{(L-1)} \times x_{j}^{(L-1)}$                      \\ \hline
Act            & $x_i^{(L)} = f(x_i^{L-1})$                      \\ \hline
C              & $x_{ij}^{(L)} = \sum_{a=0}^{s_q-1} \sum_{b=0}^{s_q-1} W_{ab}^{(L-1)} \times x_{(i\cdot s_t+a)(j\cdot s_t+b)}^{L-1}$                      \\ \hline
MeP            & $x_{ij}^{(L)} = \mathrm{Mean} (x_{(i+a)(j+b)}^{L-1}),\ a,b\in\{1,2,...,s_q\}$                      \\ \hline
MaP            & $x_{ij}^{(L)} = \mathrm{Max} (x_{(i+a)(j+b)}^{L-1}),\ a,b\in\{1,2,...,s_q\}$                      \\ \hline
\end{tabular}
}
\end{table}

\begin{table*}[]
\centering
\caption{Comparison of secure deep learning frameworks, their characteristics, and performance results for classifying one image from the MNIST dataset in the {\bf LAN} setting.}
\label{tab:cryptonets}
\resizebox{\textwidth}{!}{
\begin{tabular}{|l|l|c|r|r|r|r|r|r|r|}
\hline
\multirow{2}{*}{Framework} & \multirow{2}{*}{Methodology} & \multirow{2}{*}{\begin{tabular}[c]{@{}c@{}}Non-linear Activation \\ and Pooling Functions\end{tabular}}& \multicolumn{3}{c|}{Classification Timing (s)} & \multicolumn{3}{c|}{\begin{tabular}[c]{@{}l@{}}Communication (MB)\end{tabular}} & \multicolumn{1}{c|}{\multirow{2}{*}{\begin{tabular}[c]{@{}l@{}}Classification \\ Accuracy\end{tabular}}} \\ \cline{4-9}
                           &                              &                                                                                                        & Offline       & Online       & Total      & Offline                         & Online                        & Total                        &                                                                                                                                                                              \\ \hline
{\bf Microsoft CryptoNets}~\cite{dowlin2016cryptonets}       & Leveled HE                   & \xmark                                                                                                      & -             & -        & 297.5      & -                               & -                         & 372.2                        & 98.95\%                                                                                                                                                                  \\ \hline
{\bf DeepSecure}~\cite{rouhani2017deepsecure}                 & GC                           & \cmark                                                                                                     & -             & -         & 9.67       & -                               & -                           & 791                          & 99\%                                                                                                                                                                    \\ \hline
{\bf SecureML}~\cite{mohassel2017secureml}                   & Linearly HE, GC, SS                    & \xmark                                                                                                     & 4.70           & 0.18         & 4.88       & -                               & -                             & -                            & 93.1\%                                                                                                                                                                       \\ \hline
{\bf MiniONN} (Sqr Act.)~\cite{liuoblivious}        & Additively HE, GC, SS                            & \xmark                                                                                                     & 0.90           & 0.14         & 1.04       & 3.8                             & 12                            & 15.8                         & 97.6\%                                                                                                                                                                       \\ \hline
{\bf MiniONN} (ReLu + Pooling)~\cite{liuoblivious}   & Additively HE, GC, SS                             & \cmark                                                                                                    & 3.58          & 5.74         & 9.32       & 20.9                            & 636.6                         & 657.5                        & 99\%                                                                                                                                                                         \\ \hline
{\bf EzPC}~\cite{EzPC}                  & GC, Additive SS         & \cmark                                                                                                     & -          & -         & 5.1       & -                               & -                             & 501                         & 99\%                                                                                                                                                                   \\ \hline
{\bf Chameleon (This Work)}                  & GC, GMW, Additive SS         & \cmark                                                                                                     & 1.25          & 0.99         & 2.24       & 5.4                               & 5.1                             & 10.5                         & 99\%                                                                                                                                                                   \\ \hline
\end{tabular}}
\end{table*}

{\bf Comparison with Previous Works (MNIST Dataset).} A comparison of recent works is given in \tab{tab:cryptonets} and described next. 
We use the MNIST dataset~\cite{MNISTDATA} (same as Microsoft CryptoNets) containing \numprint{60000} images of hand-written digits. Each image is represented as $28\times28$ pixels with values between 0 and 255 in gray scale. We also train the same NN architecture using the Keras library~\cite{chollet2015} running on top of TensorFlow~\cite{tensorflow2015} using \numprint{50000} images. We achieve a similar test accuracy of $\sim 99\%$ examined over \numprint{10000} test images. The architecture of the trained CNN is depicted in \fig{fig:arch} and composed of (i) $C$ layer with a kernel of size $5\times5$, stride $2$, and map count of $5$. (ii) $Act$ layer with ReLu as the activation function. (iii) A $FC^{980\times100}$ layer. (iv) Another ReLu $Act$ layer, and (v) a $FC^{100\times10}$ layer. 
The lower bar in \fig{fig:arch} shows the different layers of the CNN while the upper bar depicts the corresponding protocol that executes each part of the CNN.

The output of the last layer is a vector of ten numbers where each number represents the probability of the image being each digit (0-9). We extract the maximum value and output it as the classification result. 
The trained CNN is the server's input and the client's input is the image that is going to be classified. More precisely, the trained model consists of the kernels' values and weights (matrices) of the FC layers. The output of the secure computation is the classification (inference) label.

The performance results compared with Microsoft CryptoNets and most recent works are provided in \tab{tab:cryptonets}.  
We report our run-time as Offline/Online/Total. As can be seen, \sys{} is 133x faster compared to the customized solution based on homomorphic encryption of CryptoNets~\cite{dowlin2016cryptonets}. They performed the experiments on a similar machine (Intel Xeon ES-1620 CPU @ \SI{3.5}{\giga\hertz} with \SI{16}{\giga\byte} of RAM). Please note that in CryptoNets~\cite{dowlin2016cryptonets}, numbers are represented with 5 to 10 bit precision while in \sys{}, all numbers are represented as 64 bit numbers. While the precision does not considerably change the accuracy for the MNIST dataset, it might significantly reduce the accuracy results for other datasets. In addition, the CryptoNets framework neither supports non-linear activation nor pooling functions.
%
However, it is worth-mentioning that CryptoNets can process a batch of images of size \numprint{8192} with no additional costs. 
Therefore, the CryptoNets framework can process up to \numprint{51739} predictions per hour.
Nonetheless, it is necessary that the system batches a large number of images and processes them together. This, in turn, might reduce the throughput of the network significantly. 
\CHANGED{
A recent solution based on leveled homomorphic encryption is called CryptoDL~\cite{hesamifard2017cryptodl}.
In CryptoDL, several activation functions are approximated using low-degree polynomials and mean-pooling is used as a replacement for max-pooling.
The authors state up to \numprint{163840} predictions per hour for the same batch size as in CryptoNets.
For a single instance, CryptoDL incurs the same computation and communication costs as for one batch.
Also, note that in Chameleon one can implement and evaluate virtually any activation and pooling function.
}

The DeepSecure framework~\cite{rouhani2017deepsecure} is a GC-based framework for secure Deep Learning inference. DeepSecure also proposes data-level and network-level preprocessing step before the secure computation protocol. They report a classification run-time of \SI{9.67}{\second} to classify images from the MNIST dataset using a CNN similar to CryptoNets. They utilize non-linear activation and pooling functions. \sys{} is 4.3x faster and requires 75x less communication compared to DeepSecure when running an identical CNN.

SecureML~\cite{mohassel2017secureml} is a framework for privacy-preserving machine learning. Similar to CryptoNets, SecureML focuses on linear activation functions. The MiniONN~\cite{liuoblivious} framework reduces the classification latency on an identical network from \SI{4.88}{\second} to \SI{1.04}{\second} using similar linear activation functions. MiniONN also supports non-linear activation functions and max-pooling. They report a classification latency of \SI{9.32}{\second} while successfully classifying MNIST images with 99\% accuracy. For a similar accuracy and network, \sys{} has 4.2x lower latency and requires 63x less communication.

\CHANGED{
For the evaluation of the very recent EzPC framework \cite{EzPC}, the authors implement the CNN from MiniONN in a high-level language.
The EzPC compiler translates this implementation to standard ABY input while automatically inserting conversions between GC and A-SS.
This results in a total run-time of \SI{5.1}{\second} for classifying one image.
However, note that Chameleon requires 48x less communication.}

\tab{tab:cryptonets} shows that the total run-time of the end-to-end execution of \sys{} for a single image is only \SI{2.24}{\second}. However, \sys{} can easily be scaled up to classify multiple images at the same time using a CNN with non-linear activation and pooling functions. For a batch size of~100, our framework requires only \SI{0.18}{\second} processing time and \SI{10.5}{\mega\byte} communication per image providing up to \numprint{20000} predictions per hour in the LAN setting.
\tab{tab:MNIST_WAN} furthermore shows the required run-times and communication for different batch sizes in a WAN setting where we restrict the bandwidth to \SI{100}{\mega\bit\per\second} with a round-trip time of \SI{100}{\milli\second}.
\CHANGED{In the WAN setting, we replace all GMW protocol invocations with the GC protocol to benefit from its constant round property.}

\begin{table}[]
\centering
\caption{Classification time (in seconds) and communication costs (in megabytes) of \sys{} for different batch sizes of the MNIST dataset in the {\bf WAN}  setting (\SI{100}{\mega\bit\per\second} bandwidth, \SI{100}{\milli\second} round-trip time).}
\label{tab:MNIST_WAN}
\resizebox{\columnwidth}{!}{
\begin{tabular}{r|r|r|r|r|r|r|}
\cline{2-7}
									&	\multicolumn{3}{c|}{Classification Time (s)} & \multicolumn{3}{c|}{Communication (MB)} \\ \hline
\multicolumn{1}{|r|}{Batch Size}	&	Offline	& Online	&	Total	&	Offline	&	Online	&	Total	\\ \hline
\multicolumn{1}{|r|}{1}				&	4.03	&	2.85	&	6.88	&	7.8		&	5.1		&	12.9	\\ \hline
\multicolumn{1}{|r|}{10}			&	10.00	&	10.65	&	20.65	&	78.4	&	50.5	&	128.9	\\ \hline
\multicolumn{1}{|r|}{100}			&	69.38	&	84.09	&	153.47	&	784.1	&	505.3	&	1289.4	\\ \hline
\end{tabular}
}
\end{table}


{\bf Comparison with Previous Works (CIFAR-10 Dataset).}
In accordance with previous works, we also evaluate our framework by running a CNN for classifying images from the CIFAR-10 dataset \cite{CIFAR10}.
The CIFAR-10 dataset comprises \numprint{60000} color images with a resolution of 32 x 32 pixels.
We implement and train a CNN with the same architecture as given in Fig.\ 13 in \cite{liuoblivious}, which achieves \SI{81.61}{\percent} accuracy.
Compared to the CNN used for classifying MNIST images, the architecture of this CNN is more sophisticated: in total there are 7 convolution layers, 7 ReLu activation layers, 2 mean-pooling layers, and one fully connected layer.
In \fig{fig:cifar10_cnn} we give the architecture of the CNN implemented and trained for classifying images from the CIFAR-10 dataset.
The architecture is the same as the one in Fig.\ 13 in \cite{liuoblivious}.
We also list the protocols used to execute each layer of the CNN in \sys{} and the necessary protocol conversions.
We report the performance results when classifying one image in \tab{tab:CIFAR-10}.
Compared to MiniONN \cite{liuoblivious}, the total run-time is reduced by factor 10.3x.
The more recent EzPC framework \cite{EzPC} is still by factor 5x slower than our solution and requires 15x more communication.

\begin{figure}[H]
\centering
\resizebox{\columnwidth}{!}{
\begin{tabular}{|l|L{7cm}|c|}
\hline
Layer					& Description						& Protocol	\\ \hline\hline

Convolution				& Input image $3 \times 32 \times 32$, window size $3 \times 3$, stride $(1, 1)$, pad $(1, 1)$, number of output channels 64: $\mathbb{R}^{64 \times 1024} \leftarrow \mathbb{R}^{64 \times 27} \cdot \mathbb{R}^{27 \times 1024}$.	& A-SS	\\ \hline

\multicolumn{3}{|c|}{\textit{A2GMW}} \\ \hline

ReLu Activation			& Computes ReLu for each input.	& GMW	\\ \hline

\multicolumn{3}{|c|}{\textit{GMW2A}} \\ \hline

Convolution				& Window size $3 \times 3$, stride $(1, 1)$, pad $(1, 1)$, number of output channels 64: $\mathbb{R}^{64 \times 1024} \leftarrow \mathbb{R}^{64 \times 576} \cdot \mathbb{R}^{576 \times 1024}$.	& A-SS	\\ \hline

\multicolumn{3}{|c|}{\textit{A2GMW}} \\ \hline

ReLu Activation			& Computes ReLu for each input.	& GMW	\\ \hline

\multicolumn{3}{|c|}{\textit{GMW2A}} \\ \hline

Mean Pooling			& Window size $1 \times 2 \times 2$, outputs $\mathbb{R}^{64 \times 16 \times 16}$.	& A-SS	\\ \hline

Convolution				& Window size $3 \times 3$, stride $(1, 1)$, pad $(1, 1)$, number of output channels 64: $\mathbb{R}^{64 \times 256} \leftarrow \mathbb{R}^{64 \times 576} \cdot \mathbb{R}^{576 \times 256}$.	& A-SS	\\ \hline

\multicolumn{3}{|c|}{\textit{A2GMW}} \\ \hline

ReLu Activation			& Computes ReLu for each input.	& GMW	\\ \hline

\multicolumn{3}{|c|}{\textit{GMW2A}} \\ \hline

Convolution				& Window size $3 \times 3$, stride $(1, 1)$, pad $(1, 1)$, number of output channels 64: $\mathbb{R}^{64 \times 256} \leftarrow \mathbb{R}^{64 \times 576} \cdot \mathbb{R}^{576 \times 256}$.	& A-SS	\\ \hline

\multicolumn{3}{|c|}{\textit{A2GMW}} \\ \hline

ReLu Activation			& Computes ReLu for each input.	& GMW	\\ \hline

\multicolumn{3}{|c|}{\textit{GMW2A}} \\ \hline

Mean Pooling			& Window size $1 \times 2 \times 2$, outputs $\mathbb{R}^{64 \times 16 \times 16}$.	& A-SS	\\ \hline

Convolution				& Window size $3 \times 3$, stride $(1, 1)$, pad $(1, 1)$, number of output channels 64: $\mathbb{R}^{64 \times 64} \leftarrow \mathbb{R}^{64 \times 576} \cdot \mathbb{R}^{576 \times 64}$.	& A-SS	\\ \hline

\multicolumn{3}{|c|}{\textit{A2GMW}} \\ \hline

ReLu Activation			& Computes ReLu for each input.	& GMW	\\ \hline

\multicolumn{3}{|c|}{\textit{GMW2A}} \\ \hline

Convolution				& Window size $1 \times 1$, stride $(1, 1)$, number of output channels 64: $\mathbb{R}^{64 \times 64} \leftarrow \mathbb{R}^{64 \times 64} \cdot \mathbb{R}^{64 \times 64}$.	& A-SS	\\ \hline

\multicolumn{3}{|c|}{\textit{A2GMW}} \\ \hline

ReLu Activation			& Computes ReLu for each input.	& GMW	\\ \hline

\multicolumn{3}{|c|}{\textit{GMW2A}} \\ \hline

Convolution				& Window size $1 \times 1$, stride $(1, 1)$, number of output channels 16: $\mathbb{R}^{16 \times 64} \leftarrow \mathbb{R}^{16 \times 64} \cdot \mathbb{R}^{64 \times 64}$.	& A-SS	\\ \hline

\multicolumn{3}{|c|}{\textit{A2GMW}} \\ \hline

ReLu Activation			& Computes ReLu for each input.	& GMW	\\ \hline

\multicolumn{3}{|c|}{\textit{GMW2A}} \\ \hline

Fully Connected Layer	& Fully connects the incoming 1024 nodes to the outgoing 10 nodes: $\mathbb{R}^{10 \times 1} \leftarrow \mathbb{R}^{10 \times 1024} \cdot \mathbb{R}^{1024 \times 1}$.	& A-SS	\\ \hline

\multicolumn{3}{|c|}{\textit{A2GC}} \\ \hline

Arg Max					& Extracts the label of the class with the highest probability.		& GC	\\ \hline

\end{tabular}
}

\caption{\label{fig:cifar10_cnn}{The architecture of the CNN trained from the CIFAR-10 dataset (taken from \cite{liuoblivious}) and the protocols used to execute each layer in \sys{}, including the necessary protocol conversions.}}
\end{figure}

\begin{table}[]
\centering
\caption{Classification time (in seconds) and communication costs (in gigabytes) of secure deep learning frameworks for one image from the CIFAR-10 dataset in the LAN setting.}
\label{tab:CIFAR-10}
\resizebox{\columnwidth}{!}{
\begin{tabular}{|l|r|r|r|r|r|r|}
\hline
\multirow{2}{*}{Framework}	&	\multicolumn{3}{c|}{Classification Time (s)} & \multicolumn{3}{c|}{Communication (GB)} \\\cline{2-7}
													& Offline	& Online	& Total	& Offline	& Online	& Total	\\ \hline
{\bf MiniONN} \cite{liuoblivious}	& 472		& 72		& 544	& 6.23		& 3.05		& 9.28	\\ \hline
{\bf EzPC} \cite{EzPC}				& -			& -			& 265.6	& -			& -			& 40.63	\\ \hline
{\bf Chameleon (This Work)}			& 22.97		& 29.7		& 52.67	& 1.21		& 1.44		& 2.65	\\ \hline
\end{tabular}
}
\end{table}

{\bf Further Related Works.}
One of the earliest solutions for obliviously evaluating a neural network was proposed by Orlandi et al.~\cite{OPB07}. They suggest adding fake neurons to the hidden layers in the original network and evaluating the network using HE.
Chabanne et al.~\cite{chabanne2017privacy} also approximate the ReLu non-linear activation function using low-degree polynomials and provide a normalization layer prior to the activation layer. However, they do not report experimental results. 
Sadeghi and Schneider proposed to utilize universal circuits to securely evaluate neural networks and fully hide their structure~\cite{SS09}.
Privacy-preserving classification of electrocardiogram (ECG) signals using neural networks has been addressed in~\cite{BFLSS11}.
The recent work of Shokri and Shmatikov~\cite{shokri2015privacy} is a Differential Privacy (DP) based approach for the distributed training of a Neural Network and they do not provide secure DNN or CNN inference. Due to the added noise in DP, any attempt to implement secure inference suffers from a significant reduction in accuracy of the prediction. Phong et al.~\cite{leprivacy} propose a mechanism for privacy-preserving deep learning based on additively homomorphic encryption. They do not consider secure deep learning inference (classification). There are also limitations of deep learning when an adversary can craft malicious inputs in the training phase~\cite{AdversarialDL}. Moreover, deep learning can be used to break semantic image CAPTCHAs~\cite{DLCAPTCHA}.

\subsection{Support Vector Machines (SVMs)}\label{ssec:SVM}
One of the most frequently used classification tools in machine learning and data mining is the Support Vector Machine (SVM). An SVM is a supervised learning method in which the model is created based on labeled training data. The result of the training phase is a non-probabilistic {\it binary} classifier. The model can then be used to classify input data ${\bf x}$ which is a $d$-dimensional vector. In \sys{}, we are interested in a scenario where the server holds an already trained SVM model and the user holds the query ${\bf x}$. Our goal is to classify the user's query without disclosing the user's input to the server or the server's model to the user. 

\begin{figure}[ht]
\centering
\includegraphics[width=0.55\columnwidth]{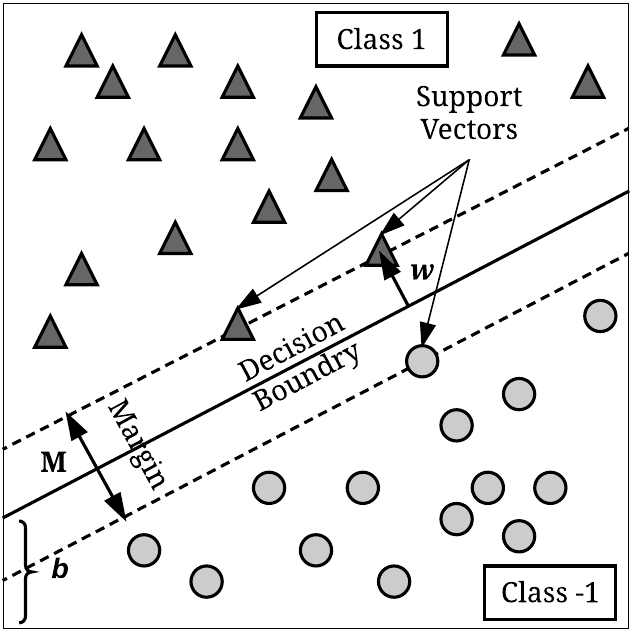}
\caption{\label{fig:svm}{Classification using Support Vector Machine (SVM).}}
\end{figure}

The training data, composed of $N$ $d$-dimensional vectors, can be viewed as $N$ points in a $d$-dimensional space. Each point $i$ is labeled as either ${\bf y_i}\in\{-1,1\}$, indicating which class the data point belongs to. If the two classes are linearly separable, a $(d-1)$-dimensional hyperplane that separates these two classes can be used to classify future queries. A new query point can be labeled based on which side of the hyperplane it resides on. The hyperplane is called {\it decision boundary}. While there can be infinitely many such hyperplanes, a hyperplane is chosen that maximizes the margin between the two classes. That is, a hyperplane is chosen such that the distance between the nearest point of each class to the hyperplane is maximized. Those training points that reside on the margin are called {\it support vectors}. This hyperplane is chosen to achieve the highest classification accuracy. \fig{fig:svm} illustrates an example for a two-dimensional space. The optimal hyperplane can be represented using a vector ${\bf w}$ and a distance from the origin {\bf b}. Therefore, the optimization task can be formulated as: 
$$\textrm{minimize}\ \lVert{{\bf w}}\rVert \ \textrm{s.t.}\ {\bf y_i}({\bf w}\cdot{\bf x_i}-{\bf b})\geqslant 1,\  i=1,2,...,N$$
The size of the margin equals ${\bf M}=\frac{2}{\lVert{{\bf w}}\rVert}$. This approach is called hard-margin SVM.

An extension of the hard-margin SVM, called a soft-margin SVM, is used for scenarios where the two classes are not linearly separable. In this case, the hinge lost function is used to penalize if the training sample is residing on the wrong side of the classification boundary. As a result, the optimization task is modified to:
$$\frac{1}{N}\sum^{N}_{i=1} \max\ (0,1-{\bf y_i}({\bf w}\cdot{\bf x_i}-{\bf b}))+\lambda \lVert{{\bf w}}\rVert^2$$
where $\lambda$ is a parameter for the tradeoff between the size of the margin and the number of points that lie on the correct side of the boundary.

For both soft-margin and hard-margin SVMs, the performed classification task is similar. The output label of the user's query is computed as:
$$\mathsf{label} \in\{-1,1\}= \textrm{sign} ({\bf w}\cdot{\bf x}-{\bf b})$$

We run our experiments using the same setup described in \sect{sec:app}. The results of the experiments are provided in \tab{svm-tab} for feature vector sizes of \numprint{10}, \numprint{100}, and \numprint{1000}.

\begin{table}[]
\centering
\caption{Classification time (in seconds) and communication costs (in kilobytes) of \sys{} using SVM models for different feature sizes in the LAN setting.}
\label{svm-tab}
\resizebox{\columnwidth}{!}{
\begin{tabular}{r|r|r|r|r|r|r|}
\cline{2-7}
                                   & \multicolumn{3}{l|}{Classification Time (ms)} & \multicolumn{3}{l|}{Communication (kB)} \\ \hline
\multicolumn{1}{|c|}{Feature Size} & Offline         & Online        & Total        & Offline      & Online      & Total      \\ \hline
\multicolumn{1}{|r|}{10}           & 8.91            & 0.97          & 9.88         & 3.2          & 3.3		 & 6.5        \\ \hline
\multicolumn{1}{|r|}{100}          & 9.49            & 0.99          & 10.48        & 3.9          & 4.7         & 8.7        \\ \hline
\multicolumn{1}{|r|}{1000}         & 10.28           & 1.14          & 11.42        & 11.1         & 19.1        & 30.3       \\ \hline
\end{tabular}
}
\end{table}

\begin{table*}[h]
\centering
\caption{{\bf Run-Times} (in milliseconds unless stated otherwise) for different atomic operations and comparison with prior art. Each experiment is performed for \numprint{1000} operations on 32-bit numbers in parallel. The detailed performance results for ABY~\cite{aby} are provided for three different modes of operation: GC, GMW, and Additive. Minimum values marked in {\bf bold}.}
\label{tab:atomic_runtime}
\resizebox{\textwidth}{!}{
\begin{tabular}{l|r|r|r|r|r|r|r|r|r|r|}
\cline{2-11}
	&	\multicolumn{1}{c|}{{\bf TinyGarble}~\cite{songhori2015tinygarble}}	&	\multicolumn{2}{c|}{{\bf ABY-GC}~\cite{aby}}	&	\multicolumn{2}{c|}{{\bf ABY-GMW}~\cite{aby}}	&	\multicolumn{2}{c|}{{\bf ABY-A}~\cite{aby}}	&	\multicolumn{1}{c|}{{\bf Sharemind}~\cite{bogdanov2008sharemind}}	&	\multicolumn{2}{c|}{{\bf Chameleon}}	\\ \hline
\multicolumn{1}{|l|}{{\bf Op}}		&	\multicolumn{1}{C{3.75cm}|}{Online} & \multicolumn{1}{C{1.5cm}|}{Offline} &  \multicolumn{1}{C{1.5cm}|}{Online} & \multicolumn{1}{C{1.5cm}|}{Offline} & \multicolumn{1}{C{1.5cm}|}{Online} & \multicolumn{1}{C{1.5cm}|}{Offline} & \multicolumn{1}{C{1.5cm}|}{Online} & \multicolumn{1}{C{3.75cm}|}{Online} & \multicolumn{1}{C{1cm}|}{Offline} & \multicolumn{1}{C{1cm}|}{Online} \\ \hline
\multicolumn{1}{|l|}{{\bf ADD}}		&	1.57 s			&	11.71		&	2.73		&	25.78		& 	4.73		&	{\bf 0.00}					&	{\bf 0.00}					& 1 $\mu$s		&	{\bf 0.00}	&	{\bf 0.00}	\\ \hline
\multicolumn{1}{|l|}{{\bf MULT}}	&	2.31 s			&	423.82		&	112.29		&	174.52		&	14.25		&	10.46						&	0.59						& 17			&	{\bf 4.24}	&	{\bf 0.13}	\\ \hline
\multicolumn{1}{|l|}{{\bf XOR}}		&	{\bf 0.00}		&	{\bf 0.00}	&	{\bf 0.00}	&	{\bf 0.00}	&	{\bf 0.00}	&	\cellcolor[HTML]{dbdbdb}	&	\cellcolor[HTML]{dbdbdb}	& 1 $\mu$s		&	{\bf 0.00}	&	{\bf 0.00}	\\ \hline
\multicolumn{1}{|l|}{{\bf AND}}		&	1.58 s 			&	11.83		&	2.34		&	9.27		&	{\bf 0.52}	&	\cellcolor[HTML]{dbdbdb}	&	\cellcolor[HTML]{dbdbdb}	& 17			&	{\bf 1.50}	&	0.56		\\ \hline
\multicolumn{1}{|l|}{{\bf CMP}}		&	1.57 s			&	11.90		&	2.63		&	17.39		&	1.63		&	\cellcolor[HTML]{dbdbdb}	&	\cellcolor[HTML]{dbdbdb}	& 2.5 s			&	{\bf 2.46}	&	{\bf 1.48}	\\ \hline
\multicolumn{1}{|l|}{{\bf EQ}}		&	1.56 s			&	11.60		&	2.42 		&	9.11		&	1.15		&	\cellcolor[HTML]{dbdbdb}	&	\cellcolor[HTML]{dbdbdb}	& 5 s			&	{\bf 1.54}	&	{\bf 1.09}	\\ \hline
\multicolumn{1}{|l|}{{\bf MUX}}		&	1.59 s			&	11.91		&	2.49		&	{\bf 1.06}	&	0.68		&	\cellcolor[HTML]{dbdbdb}	&	\cellcolor[HTML]{dbdbdb}	& 34			&	1.52		&	{\bf 0.63}	\\ \hline
\end{tabular}}
\end{table*}

\begin{table*}[ht]
\centering
\caption{{\bf Communication} (in kilobytes unless stated otherwise) for different atomic operations and comparison with prior art. Each experiment is performed for \numprint{1000} operations on 32-bit numbers in parallel. The detailed performance results of the ABY framework~\cite{aby} is provided for three modes of operation: GC, GMW, and Additive. Minimum values marked in {\bf bold}.}
\label{tab:atomic_communication}
\resizebox{\textwidth}{!}{
\begin{tabular}{l|r|r|r|r|r|r|r|r|r|r|}
\cline{2-11}
                           & \multicolumn{1}{c|}{{\bf TinyGarble}~\cite{songhori2015tinygarble}}                                                        & \multicolumn{2}{c|}{{\bf ABY-GC}~\cite{aby}}                                           & \multicolumn{2}{c|}{{\bf ABY-GMW}~\cite{aby}}                                          & \multicolumn{2}{c|}{{\bf ABY-A}~\cite{aby}}                                            & \multicolumn{1}{c|}{{\bf Sharemind}~\cite{bogdanov2008sharemind}}                                        & \multicolumn{2}{c|}{{\bf Chameleon}}                                        \\ \hline
\multicolumn{1}{|l|}{{\bf Op}}		&	\multicolumn{1}{c|}{Total} & \multicolumn{1}{C{1.5cm}|}{Offline} &  \multicolumn{1}{C{1.5cm}|}{Online} & \multicolumn{1}{C{1.5cm}|}{Offline} & \multicolumn{1}{C{1.5cm}|}{Online} & \multicolumn{1}{C{1.5cm}|}{Offline} & \multicolumn{1}{C{1.5cm}|}{Online} & \multicolumn{1}{c|}{Total} & \multicolumn{1}{C{1cm}|}{Offline} & \multicolumn{1}{C{1cm}|}{Online} \\ \hline
\multicolumn{1}{|l|}{{\bf ADD}}   & 7936			& 992					&         {\bf 0}              & 3593 	&	76		&	{\bf 0}						&	{\bf 0}						&	{\bf 0}						&	{\bf 0}		& {\bf 0}	\\ \hline
\multicolumn{1}{|l|}{{\bf MULT}}  & 318 K			& 47649 				&         {\bf 0}	           & 37900 	&	840		&	1280               			&	16							&	192							&	{\bf 8}		& 16		\\ \hline
\multicolumn{1}{|l|}{{\bf XOR}}   & {\bf 0}			& {\bf 0}				&         {\bf 0}              & {\bf 0}&	{\bf 0}	&	\cellcolor[HTML]{dbdbdb}	&	\cellcolor[HTML]{dbdbdb}  	&	{\bf 0}						&	{\bf 0}		& {\bf 0}	\\ \hline
\multicolumn{1}{|l|}{{\bf AND}}   & 8192 			& 1024					&         {\bf 0}              & 1028 	&	16		&	\cellcolor[HTML]{dbdbdb}	&	\cellcolor[HTML]{dbdbdb} 	&	192							&	{\bf 12}	& 8			\\ \hline
\multicolumn{1}{|l|}{{\bf CMP}}   &	8192			& 1024					&         {\bf 0}              & 2851 	&	45		&	\cellcolor[HTML]{dbdbdb}	&   \cellcolor[HTML]{dbdbdb}  	&	\cellcolor[HTML]{dbdbdb}    &	{\bf 23}	& 33		\\ \hline
\multicolumn{1}{|l|}{{\bf EQ}}    &	7936			& 992					&         {\bf 0}              & 995	&	16		&	\cellcolor[HTML]{dbdbdb}	&   \cellcolor[HTML]{dbdbdb}  	&	\cellcolor[HTML]{dbdbdb}    &	{\bf 8}   	& 12		\\ \hline
\multicolumn{1}{|l|}{{\bf MUX}}   & 8192			& 1024					&         {\bf 0}              & 33	 	&	8		&	\cellcolor[HTML]{dbdbdb}	&   \cellcolor[HTML]{dbdbdb}  	&	384							&	{\bf 8}		& 4			\\ \hline
\end{tabular}}
\end{table*}

\textbf{Comparison with Previous Works.}
\CHANGED{
Makri et al.~\cite{PICS} present PICS, a private image classification system based on SVM learning.
They evaluate their implementation in SPDZ \cite{damgaard2012multiparty} with two computation nodes.
For one binary classification with 20 features, they report \SI{145}{\second}/\SI{30}{\milli\second} offline/online run-time.
Although in a different security and computational model, \sys{} performs the same task four orders of magnitude faster.
}
Bos et al.~\cite{BPTG15} study privacy-preserving classification based on hyperplane decision, Naive Bayes, and decision trees using homomorphic encryption. For a credit approval dataset with 47 features, they report a run-time of \SI{217}{\milli\second} and \SI{40}{\kilo\byte} of communication, whereas, \sys{} can securely classify a query with \numprint{1000} features in only \SI{11.42}{\milli\second} with \SI{30.3}{\kilo\byte} of communication. 
Rahulamathavan et al.~\cite{RPVCR14} also design a solution based on homomorphic encryption for binary as well as multi-class classification based on SVMs. In the case of binary classification, for a dataset with 9 features, they report \SI{7.71}{\second} execution time and \SI{1.4}{\mega\byte} communication. In contrast, for the same task, \sys{} requires less than \SI{10}{\milli\second} execution time and \SI{6.5}{\kilo\byte} of communication. 
Laur et al.~\cite{LLM06} provide privacy-preserving training algorithms based on general kernel methods. They also study privacy-preserving classification based on SVMs but they do not report any benchmark results. 
Vaidya et al. \cite{vaidya2008privacy} propose a method to train an SVM model where the training data is distributed among multiple parties. This scenario is different than ours where we are interested in the SVM-based classification. As a proof-of-concept, we have focused on SVM models for linear decision boundaries. However, \sys{} can be used for non-linear decision boundaries as well.

\section{Benchmarks of Atomic Operations}\label{sec:eval}
We benchmark different atomic operations of \sys{} and compare them with three prior art frameworks: TinyGarble~\cite{songhori2015tinygarble}, ABY~\cite{aby}, and Sharemind~\cite{bogdanov2008sharemind}.
The result for ABY is reported for three different scenarios: GC-only, GMW-only, and Additive SS-only.
For TinyGarble, ABY, and \sys{} we run the frameworks ourselves.
The benchmarking environment remains the same as described in \sect{sec:app}.
Unlike TinyGarble and ABY, Sharemind lacks built-in atomic benchmarks and is a commercial product that requires contracting even for academic purposes.
Thus, we give the results from the original paper~\cite{bogdanov2008sharemind} and justify why Chameleon performs better on equal hardware.

We do not include WAN benchmarks of atomic operations for the following reason: Due to higher latency, GC-based circuit evaluation with constant rounds is preferred instead of GMW for binary operations.
However, since the atomic benchmarks do not measure input sharing (for which GC uses STP-aided OT generation), no difference is visible to prior art.


{\bf Evaluation Results.}
The detailed run-times and communication costs for arithmetic and binary operations are given in \tab{tab:atomic_runtime} and in \tab{tab:atomic_communication}, respectively.
The highlighted area for ABY-A in both tables reflects that ABY does not perform these operations in additive secret sharing. The highlighted area in \tab{tab:atomic_communication} for Sharemind indicates that the corresponding information is not reported in the original paper.
\tab{tab:conversion_runtime} additionally shows the run-times for conversions between different sharings.\footnote{The required \emph{communication} for conversion operations equals ABY \cite{aby} since STP-aided OT generation does not reduce the amount of communication (cf.\ \tab{tab:setup}).}
All reported run-times are the average of 10 executions with less than 15\% variance.

\begin{table}[h]
\centering
\caption{Run-Times (in milliseconds) for conversion operations and comparison with prior art. Each experiment is performed for \numprint{1000} operations on 32-bit numbers in parallel. Minimum values marked in {\bf bold}.}
\label{tab:conversion_runtime}
\scalebox{0.9}{
\begin{tabular}{l|r|r|r|r|r|r|r|r|}
\cline{2-5}	& \multicolumn{2}{c|}{{\bf ABY}~\cite{aby}}	& \multicolumn{2}{c|}{{\bf Chameleon}}	\\ \hline
\multicolumn{1}{|l|}{{\bf Op}}	& \multicolumn{1}{c|}{Offline}& \multicolumn{1}{c|}{Online} & \multicolumn{1}{c|}{Offline} & \multicolumn{1}{c|}{Online}	\\ \hline
\multicolumn{1}{|l|}{{\bf GC2GMW}}	& {\bf 0.00}	& {\bf 0.00}&	{\bf 0.00}		&	{\bf 0.00}		\\ \hline
\multicolumn{1}{|l|}{{\bf GMW2A}}	& 9.47			& 2.44		&	{\bf 3.45}		&	{\bf 2.33}		\\ \hline
\multicolumn{1}{|l|}{{\bf GMW2GC}}	& 17.05			& 1.30		&	{\bf 13.24}		&	{\bf 1.15}		\\ \hline
\multicolumn{1}{|l|}{{\bf A2GC}}	& 19.75			& 14.03		&	{\bf 15.83}		&	{\bf 12.91}		\\ \hline
\end{tabular}
}
\end{table}

As can be seen, \sys{} outperforms all state-of-the-art frameworks.
\CHANGED{
Run-times and communication for arithmetic operations in \sys{} are only given in A-SS since from the ABY results and \tab{tab:conversion_runtime} it follows that even for a single addition or multiplication operation it is worthwhile to perform a protocol conversion.
The remaining atomic operations for \sys{} are given in Boolean sharing where we observe major improvements over ABY due to our efficient B-MT precomputation.\footnote{The benchmarking methodology inherited from ABY omits input sharing, which is why no improvement for GC-based operations is measurable compared to ABY.}
Regarding conversion operations, the GMW2A, GMW2GC, and A2GC performance in \sys{} benefits from fast STP-aided OTs (cf.\ \sect{ssec:fot}).
}

Although, the experimental setup of Sharemind is computationally weaker than ours, we emphasize that \sys{} is more efficient because of the following reasons: (i) To compute each MULT operation, Sharemind requires 6 instances of the Du-Atallah protocol while our framework needs only 2. (ii)~In Sharemind, bit-level operations such as XOR/AND require a bit-extraction protocol which is computationally expensive. Please note that these costs are not reported by~\cite{bogdanov2008sharemind} and hence are not reflected in \tab{tab:atomic_runtime}. (iii) Operations such as CMP, EQ, and MUX can most efficiently be realized using GC/GMW protocols and as a result, \sys{} can perform these operations faster.
The run-times for TinyGarble include base OTs, online OTs, garbling/evaluating, and data transmission. This is why the run-time for MULT is not significantly higher than for other operations that require orders of magnitude fewer gates. However, in \sys{}, we precompute all OTs which significantly reduces the run-time.
Note that the shown run-times and communication results for \sys{} represent the worst case, namely for the party that receives additional data from the STP besides the required seeds for OT and MT generation.\footnote{An improved implementation could equally distribute computation and communication among the two parties by dividing the data sent by the STP evenly, thereby further reducing the run-times.}


{\bf Communication in the Offline Phase. }
The communication costs of the offline phase in \sys{} are compared to ABY \cite{aby} in \tab{tab:setup}.
To generate a single B-MT, \sys{} requires only a constant-size data transmission to one party and 256$\times$ less communication to the other party compared to ABY.
When generating a single A-MT, the required communication to the other party is reduced by factor 273$\times$/289$\times$/321$\times$ for a bitlength of 16/32/64, respectively.
This is a significant enhancement since in most machine learning applications, the main bottleneck is the vector/matrix multiplication, which requires a large amount of A-MTs.

\begin{table}[htb]
\centering
\caption{Communication (in bits) in the offline phase in \sys{} compared to prior art ABY~\cite{aby}.}
\label{tab:setup}
\resizebox{\columnwidth}{!}{
\begin{tabular}{l|r|r|r|}
\cline{2-4}
                           & {\bf ABY}~\cite{aby}   & {\bf \sys{}}	& {\bf Improvement}	\\ \hline
\multicolumn{1}{|l|}{{\bf OT}}   & 128      & 128 & -	\\ \hline
\multicolumn{1}{|l|}{{\bf B-MT}} & 256        & 1  & 256$\times$	\\ \hline
\multicolumn{1}{|l|}{{\bf A-MT (bitlength $\mathbf{\ell=16}$) }} & \numprint{4368}      & 16 & 273$\times$	\\ \hline
\multicolumn{1}{|l|}{{\bf A-MT (bitlength $\mathbf{\ell=32}$) }} & \numprint{9248}      & 32 & 289$\times$	\\ \hline
\multicolumn{1}{|l|}{{\bf A-MT (bitlength $\mathbf{\ell=64}$) }} & \numprint{20544}      & 64 & 321$\times$	\\ \hline
\end{tabular}}
\end{table}

\section{Related Work}\label{sec:rw}
\sys{} is essentially a two-party framework that uses a Semi-honest Third Party (STP) to generate correlated randomness in the offline phase.
In the following, we review the use of third parties in secure computation as well as other secure two-party and multi-party computation frameworks.

\textbf{Third Party-based Secure Computation.}\label{ssec:tpbsc}
Regarding the involvement of a third party in secure two-party computation, there have been several works that consider an outsourcing or \emph{server-aided} scenario, where the resources of one or more \emph{untrusted} servers are employed to achieve sub-linear work in the circuit size of a function, even workload distribution, and output fairness.
Realizing such a scenario can be done by either employing fully-homomorphic encryption (e.g., \cite{AJLTVW12}) or extending Yao's garbled circuit protocol (e.g., \cite{kmr12}).
Another important motivation for server-aided SFE is to address the issue of low-powered mobile devices, as done in \cite{cmtb13, clt14, demmler2014ad, cmtb15, mor16, cmtb16}.
Furthermore, server-aided secure computation can be used to achieve stronger security against active adversaries \cite{hs12}.

The secure computation framework of~\cite[Chapter~6]{cr} also utilizes correlated randomness. Beyond passive security and one STP, this framework also covers active security and multiple STPs.

\textbf{GC-based Frameworks.}\label{ssec:gcfrmwk} 
The first implementation of the GC protocol is Fairplay~\cite{malkhi2004fairplay} that allows users to write the program in a high-level language called Secure Function Definition Language (SFDL) which is translated into a Boolean circuit. FariplayMP~\cite{ben2008fairplaymp} is the extension of Fairplay to the multiparty setting. FastGC~\cite{huang2011faster} reduces the running time and memory requirements of the GC execution by using pipelining.
TinyGarble~\cite{songhori2015tinygarble} is one of the recent GC frameworks that proposes to generate compact and efficient Boolean circuits using industrial logic synthesis tools. TinyGarble also supports sequential circuits (cyclic graph representation of circuits) in addition to traditional combinational circuits (acyclic graph representation). Our GC engine implementation is based on TinyGarble. 
ObliVM~\cite{liu2015oblivm} provides a domain-specific programming language and secure computation framework that facilitates the development process. 
Frigate~\cite{mood2016frigate} is a validated compiler and circuit interpreter for secure computation. Also, the authors of~\cite{mood2016frigate} test and validate several secure computation compilers and report the corresponding limitations. 
PCF (Portable Circuit Format)~\cite{kreuter2013pcf} has introduced a compact representation of Boolean circuits that enables better scaling of secure computation programs. 
Authors in~\cite{kreuter2012billion} have shown the evaluation of a circuit with billion gates in the malicious model by parallelizing operations. 

\textbf{Secret Sharing-based Frameworks.}\label{ssec:sssfrmwk}
The Sharemind framework~\cite{bogdanov2008sharemind} is based on additive secret sharing over the ring $\mathbb{Z}_{2^{32}}$. The computation is performed with {\it three} nodes and is secure in the honest-but-curious adversary model where only one node can be corrupted. 
SEPIA~\cite{burkhart2010sepia} is a library for privacy-preserving aggregation of data for network security and monitoring. SEPIA is based on Shamir's secret sharing scheme where computation is performed by three (or more) privacy peers. 
VIFF (Virtual Ideal Functionality Framework)~\cite{damgaard2009asynchronous} is a framework that implements asynchronous secure computation protocols and is also based on Shamir's secret sharing.
PICCO~\cite{zhang2013picco} is a source-to-source compiler that generates secure multiparty computation protocols from functions written in the \texttt{C} language. The output of the compiler is a \texttt{C} program that runs the secure computation using linear secret sharing.
SPDZ~\cite{damgaard2012multiparty} is a secure computation protocol based on additive secret sharing that is secure against $n-1$ corrupted computation nodes in the malicious model. 
Recent work of~\cite{araki2016high, FLNW17, ABFLLNOWW17} introduces an efficient protocol for three-party secure computation. 
In general, for secret sharing-based frameworks, three (or more) computation nodes need to communicate in the online phase and in some cases, the communication is quadratic in the number of computation nodes. However, in \sys{}, the third node (STP) is not involved in the online phase which reduces the communication and running time. 

While \sys{} offers more flexibility compared to secret-sharing based frameworks, it is computationally more efficient compared to Sharemind and SEPIA. To perform each multiplication, Sharemind needs 6 instances of the Du-Atallah protocol~\cite{bogdanov2008sharemind} while \sys{} needs 1 (when one operand is shared) or 2 (in the general case where both operands are shared). In SEPIA~\cite{burkhart2010sepia}, all operations are performed modulo a prime number which is less efficient compared to modulo $2^l$ and also requires multiple multiplications for creating/reconstructing a share.

\textbf{Mixed Protocol Frameworks.}\label{ssec:sssfrmwk}
TASTY~\cite{henecka2010tasty} is a compiler that can generate mixed protocols based on GC and homomorphic encryption.
Several applications have been built that use mixed protocols, e.g., privacy-preserving ridge-regression~\cite{nikolaenko2013privacy}, matrix factorization~\cite{nikolaenko2013privacy}, iris and finger-code authentication~\cite{blanton2011secure}, and medical diagnostics~\cite{barni2009secure}.
Recently, a new framework for compiling two-party protocols called EzPC~\cite{EzPC} was presented.
EzPC uses ABY as its cryptographic back-end: a simple and easy-to-use imperative programming language is compiled to ABY input.
An interesting feature of EzPC is its \enquote{cost awareness}, i.e.\ its ability to automatically insert type conversion operations in order to minimize the total cost of the resulting protocol.
However, they claim that ABY's GC engine always provides better performance for binary operations than GMW and thus convert only between A-SS and GC.

Our framework \CHANGED{extends} the ABY framework~\cite{aby}.
Specifically, we add support for signed fixed-point numbers which is essential for almost all machine learning applications such as processing deep neural networks. In addition to combinational circuits, \sys{} also supports sequential circuits \CHANGED{by incorporating TinyGarble-methodology \cite{songhori2015tinygarble}} which provides more scalability. Our framework provides a faster online phase and a more efficient offline phase in terms of computation and communication due to the usage of an STP. Moreover, we implement a highly efficient vector dot product protocol based on correlated randomness generated by an STP. 

\textbf{Automatic Protocol Selection.}
The authors of~\cite{kerschbaum2014automatic} propose two methods, one heuristic and one based on integer programming, to find an optimal combination of two secure computation protocols, HE and GC. \CHANGED{This methodology has been applied to the ABY framework in CheapSMC \cite{DBLP:conf/dbsec/PattukKUM16}.} The current version of \sys{} does not provide automatic protocol selection. \CHANGED{However, the methods of \cite{kerschbaum2014automatic,DBLP:conf/dbsec/PattukKUM16,EzPC} can be applied in future work in order to automatically partition \sys{} programs.}

\section{Conclusion}\label{sec:conc}
We introduced \sys{}, a novel hybrid (mixed-protocol) secure computation framework based on 
ABY \cite{aby} that achieves unprecedented performance by (i) integrating sequential garbled circuits, (ii) providing an optimized vector dot product protocol for fast matrix multiplications, and (iii) employing a semi-honest third party in the offline phase for generating correlated randomness that is used for pre-computing OTs and multiplication triples.
In contrast to previous state-of-the-art frameworks, \sys{} supports signed fixed-point numbers.
We evaluated our framework on convolutional neural networks where it can process an image of hand-written digits 133x faster than the prior art Microsoft CryptoNets~\cite{dowlin2016cryptonets} and 4.2x faster than the most recent MiniONN \cite{liuoblivious}.


\ifanonymous\else
\paragraph{Acknowledgements}
This work has been co-funded by the DFG as part of project E4 within the CRC 1119 CROSSING and by the German Federal Ministry of Education and Research (BMBF) as well as by the Hessen State Ministry for Higher Education, Research and the Arts (HMWK) within CRISP.
\fi

\ifsubmission
	\bibliographystyle{ACM-Reference-Format-abbr}
\else
	\bibliographystyle{ACM-Reference-Format-abbr}
\fi
\bibliography{main_bib}

\clearpage

\end{document}